\documentclass{emulateapj}
\begin{document}


\title{The Clustering of Low Luminosity AGN}

\author{Anca Constantin and Michael S. Vogeley}
\affil{Department of Physics, Drexel University, Philadelphia, PA 19104}

\begin{abstract}

We present the first multi-parameter analysis of the spatial
clustering properties of nearby galaxies in which accretion activity
is present to different degrees.  We spectroscopically identify and
classify active galactic nuclei (AGN) from the Sloan Digital Sky
Survey (SDSS) DR2 main galaxy sample, which yields the most precise
measurement to date of AGN clustering.
Estimates of the redshift space two-point correlation function (CF)
reveal that Seyferts are clearly less clustered than normal galaxies,
while the clustering amplitude of LINERs is consistent with that of
the parent galaxy population. The similarities of the distributions of
host properties (color and concentration index) of Seyferts and LINERs
suggest that the difference in their clustering amplitudes is not
driven by the morphology-density relation.  We find that the
luminosity of the [\ion{O}{1}] emission shows the strongest influence
on AGN clustering, with low $L_{\rm [O I]}$ sources having the highest
clustering amplitude, $s_0$.  This trend is much stronger than the
previously detected dependence on $L_{\rm [O III]}$, which we confirm.
The fact that objects of given spectral types are clustered
differently seems correlated with a variety of their physical
properties including $L_{\rm [O I]}$ , $L_{\rm [O III]}$, the electron
density of the emitting gas $n_e$, and the obscuration level.  LINERs,
which exhibit high $s_0$, show the lowest luminosities and obscuration
levels, and relatively low $n_e$, suggesting that these objects harbor
black holes that are relatively massive yet weakly active or
inefficient in their accretion, probably due to the insufficiency of
their fuel supply.  Seyferts, which have low $s_0$, are very luminous
in both [\ion{O}{1}] and [\ion{O}{3}] and show large $n_e$, suggesting
that in these systems the black holes are less massive but accrete
quickly and efficiently enough to clearly dominate the ionization.
Star-forming galaxies -- the H {\sc ii}'s -- are weakly clustered;
because these systems are hosted mainly by blue, late type galaxies,
this trend can be understood as a consequence of both the
morphology-density and star formation rate-density relations.
However, the spectral properties of the H {\sc ii}'s suggest that
these systems hide in their centers, amidst large amounts of obscuring
material, black holes of generally low mass whose activity remains
relatively feeble.
Our own Milky Way may be a typical such case.

\end{abstract}


\section{Introduction}

Clustering analysis can be a useful tool to employ in investigating
the formation and evolution mechanisms of various astrophysical
objects.  In the case of AGN, the pattern of spatial clustering and
its implications for the environments of AGN provides constraints on
galaxy evolution models, and in particular on how well AGN trace the
normal, non-active galaxy distribution.  Comparisons of the AGN
environments with those of control samples of regular galaxies may
indicate differences that reveal what determines the activity in
galaxy nuclei, and even gauge its strength and time cycle, thus
yielding a phenomenological link between black hole growth and galaxy
formation and evolution processes.

Estimates of the two-point correlation function (CF) provide important
diagnostics for the black hole (BH) masses and duty cycle of fueling
in AGN.  If structure formation is hierarchical, then the rarest and most
massive virialized dark matter (DM) halos cluster the most strongly
\citep{kai84}, thus the clustering strength of a given object type
constrains the mass and number density of the halos in which they reside.
The DM halo mass can further be connected to the physical
properties of the resident galaxies and their central black holes
through empirical scaling relations: the black hole mass -- stellar velocity
dispersion $M_{\rm BH} - \sigma_*$ relation that holds for normal
galaxies \citep{geb00, fer00} as well as for the active ones
\citep{geb00b, fer01, nel04, onk04, gre05}, the correlations between
bulge properties and $M_{\rm BH}$ \citep{mag98, gra01}, and the
correlation between circular velocity and central velocity
dispersion \citep{fer02, bae03} that links dark matter halos and
supermassive BHs.  The number density of the DM halos can be used
to infer the duty-cycle, or the accretion life-time, as this is given
by the inverse ratio of this number density to that of the observed
objects \citep{mar01}.  Such predictions for the AGN's lifetime, and
its implications for the efficiency of BH formation \citep{hai00}, can
test the fundamental assumptions of semi-analytical models of AGN and galaxy
formation \citep{kau00, mat03}.

Studying the spatial clustering properties of different
spectroscopically-classified AGN species may link their ionization
mechanisms to properties of their hosts and environment.  Through
comparison of the clustering amplitudes of accreting-type sources such
as Seyferts and the starlight-ionized H {\sc ii} galaxies, it is
possible to contrast properties of the dark matter halos that host
different kinds and levels of galactic nuclear activity.  Particularly
interesting is the clustering of the ubiquitous Low Ionization
Emission-Line Regions (LINERs, e.g. Heckman 1980), whose physical
origin, whether thermal or non-thermal, remains ambiguous despite
exhaustive analysis of their properties
\citep{ho99,ho03,fil04,mao05,con05}.

Galaxy mergers are predicted to play a crucial role in triggering AGN
activity \citep{bar92, kau00, mat05}.  If galaxy nuclei become more
AGN-like when major mergers occur, then active galaxies of different
spectral signatures, corresponding to different degrees of
contributions from a non-thermal ionization process, may cluster
differently, reflecting variation in their
environments. Thus, a comparison of the clustering
properties of the active and non-active galaxies is an
important tool for testing the merger-accretion connection.

Through its wealth of both spectroscopic and photometric data, SDSS
\citep{yor00} is ideal for investigating the dependence of clustering
properties on a variety of emission characteristics.  The SDSS
observations offer the unique opportunity to examine simultaneously
the dependence of environment on the spectral classification of
emission-line galaxies, and on various other features that
characterize or even determine a detectable accretion process, in
conjunction with properties of their host galaxies.  Previous studies
with similar aims \citep{kau03, kau04, mil03, wak04} compared passive
galaxies with AGN as a general class and did not investigate the
properties of sub-populations such as Seyferts and LINERs
separately. In these studies, the optical classification of an object
as an AGN relies, especially among the narrow-line emitters, mainly on
the [\ion{N}{2}]/H$\alpha$ vs. [\ion{O}{3}]/H$\beta$ diagnostic
diagram. There are suspicions that such rather lenient definitions may
not best represent the accretion dominated population ~\citep{hao05}.
Using such definitions, the AGN population is found by \citet{mil03}
and \citet{wak04} to occupy a uniform fraction of galaxies, to be
common to a large range of galactic environments, and to follow the
distribution of the whole galaxy population, and therefore to be
unbiased tracers of mass in the universe.  However, when divided by
their ([\ion{O}{3}]) luminosities, the active nuclei show markedly
different clustering properties \citep{wak04}, different
star-formation histories in their hosts \citep{kau03}, and different
environmental preference \citep{kau04}.  Combined together, these
findings seem to suggest that Seyferts and LINERs, that represent
quite distinct distributions in $L_{\rm [O III]}$ \citep{kau03}, might
also cluster differently.  Such inferences are intriguing and thus
merit further clarification.  It is thus crucial to establish and
quantify the connections between the intrinsic physical properties of
the nuclear line-emitting regions and those of their host galaxies and
dark matter halos. We pursue these issues here by including important
spectral diagnostics that probe the astrophysics of AGN.

In this paper, we use SDSS DR2 data in an attempt to detect and
analyse differences in the clustering properties of (1) active and
non-active galaxies, (2) objects of distinct types of activity, i.e.,
accretion versus star-formation, and (3) different spectroscopic types
of AGN, e.g., Seyferts versus LINERs.  In order to carry out such
analysis with a good understanding of the possible biases introduced
by detection and spectral identification, we present in Section
~\ref{data} a systematic examination of the different AGN selection
criteria and their sensitivity to the dominance of an accreting-type
nuclear power source.  In Section ~\ref{csi} we estimate and compare
the redshift-space CF for subsamples of objects of different spectral
classes (Section ~\ref{class}), and investigate the degree to which
these measurements are influenced by the level of nuclear activity, as
traced by the [\ion{O}{3}] and [\ion{O}{1}] luminosities (Section
~\ref{csio1o3}), properties of the line-emitting gas, e.g., intrinsic
obscuration, as traced by the neutral hydrogen column densities $N_H$
and the electron density $n_e$ (Section ~\ref{ne}), and host
properties (Section ~\ref{host}).  We discuss in Section
~\ref{discussion} possible implications of our findings on the
features that distinguish Seyferts from LINERs, on the peculiar nature
of the Transition objects, that border the definition of AGN (Seyferts
or LINERs) and H {\sc ii}, and on the potentially great similarity
between the Galactic Center and a typical H {\sc ii} system. Section
~\ref{summary} summarizes the findings and conclusions of this work.


\section{The data} \label{data}

This study is based on data from the Sloan Digital Sky Survey (SDSS)
Data Release 2 \citep{aba04}.  The parent sample from which we draw
various categories of sources is a subset of the main galaxy sample
used for large-scale structure studies \citep{str02}, and includes
objects with Petrosian $r$ magnitudes $14.5 < m_r < 17.77$ after
correction for Galactic extinction based on maps of \citep{sch98}, and
with a redshift distribution that extends from $\sim0.005$ to
$\sim0.30$, with a median of $z \approx 0.1$.  Thus, the sample does
not include low redshift quasars or broad emission-line, type 1 AGN
~\citep{ric02,sch02,sch03}.  To clarify the nomenclature, what we
define as Seyferts and LINERs are only type 2 objects, where no
obvious broad component is detected, thereby leaving the narrow lines as
the prominent spectral features.

The core of the analysis we present here exploits the source
spectroscopic properties and, consequently, the source spectral type
definition.  The spectroscopic measurements we employ are drawn from a
catalog of absorption and emission line fluxes and equivalent widths
(EW) of a superset of the DR2 galaxy targets which fulfill the galaxy
sample selection criteria. These data are kindly made publicly
available by the MPA/JHU collaboration at {\tt
http://www.mpa-garching.mpg.de/SDSS/} \citep{bri04}, and are analyzed
by, e.g., \citet{kau03, kau04}.  This dataset has the advantage that
the line emission component is separated from the host galaxy light;
based on stellar population synthesis templates that span a relatively
large range in age and metallicity, the absorption-line contribution
is identified and subtracted from the total galaxy spectrum
~\citep{tre04}.  We should note here that such model-based
measurements exclude alpha-enhancement that might result in an
enhanced ``bluing'' on the evolutionary tracks that translate in a
possible enhancement in the $H\beta$ absorption index by up to
0.26\AA\ ~\citep{tho03}.  Such bias remains however negligible for the
strongly line emission systems that we consider in this study, where
we employ only $>$ 2-sigma line flux measurements (see Section
~\ref{samples}); in such samples, the minimum strength (equivalent
width) of the H$\beta$ emission is $1.5$\AA, hence, even an extreme
``bluing'' has a maximum effect of $17$\%, and should affect only
$<10$\% of the sample.

Other observed properties of the objects used in this study were
measured using the standard SDSS photometric and spectroscopic
pipelines and were obtained directly from the SDSS archive.  We employ
Petrosian magnitudes, and in calculating the absolute magnitudes from
the apparent magnitudes $m_r$ and the redshift $z$ we apply the
formula
\begin{equation}
M_r = m_r - 5~ {\rm log}_{10}(s(1+z)/1{\rm Mpc}) - 25 - K(z) + 5~{\rm
log}_{10}h + Q(z).
\end{equation}
Here $K(z)$ is the $r$-band magnitude galaxy $K$-correction relative
to the $z = 0.1$ value, as calculated by \citet{bla03b}, and made
publicly available as part of the NYU Value-Added Catalog (Blanton et
al. 2005, {\tt http://wassup.physics.nyu.edu/vagc/}), $s(1+z)$ is the
luminosity distance, with $s$ the comoving distance, and $Q(z)$
accounts for the average evolution in galaxy luminosities in the
recent past, with $Q(z) = 1.6(z - 0.1)$ \citep{bla03a,teg04}.
Throughout this paper we assume a cosmology with $\Omega = 0.3$,
$\Lambda = 0.7$, and $H_0 = 100h$ km s$^{-1}$ Mpc$^{-1}$. The
distances are quoted in $h^{-1}$ Mpc.

\begin{figure}
\plotone{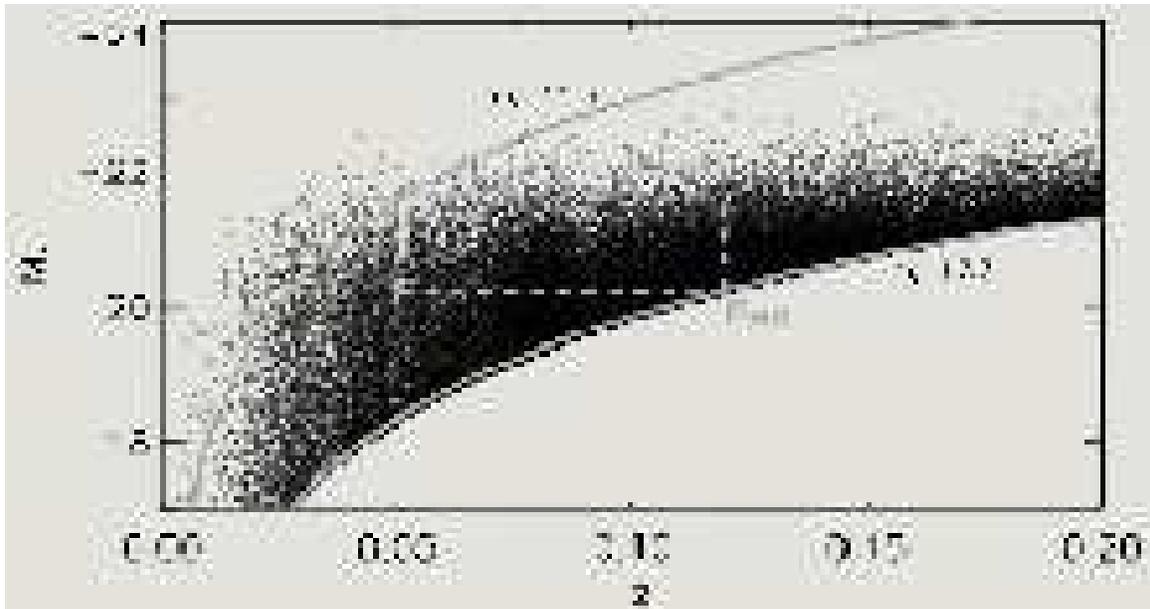}
\caption{ Volume-limited sample definition in
  redshift-absolute magnitude space.  The smooth curves delineate the
  SDSS main galaxy sample boundaries corresponding to our choice of
  apparent magnitude limits of $14.5 < m_r < 17.7$.  The rectangle
  shows the limits in absolute magnitude of the parent sample
  of galaxies used in this analysis.
\label{vol}}
\end{figure}

In this study we conduct a comparative analysis of the clustering
properties of the emission-line galaxies by employing subsamples of
galaxies, defined by absolute magnitude limits.  Figure ~\ref{vol}
illustrates the distribution of the whole galaxy sample (247,080
objects) in the redshift-absolute magnitude space.  The continuous
curves delineate the sample boundaries corresponding to our choice of
apparent magnitude limits $14.5 < m_r < 17.7$.  The bright
cutoff excludes bright galaxies that may have been shredded into many
pieces, particularly in early SDSS photometry, and/or galaxies that
have been excluded from spectroscopic target selection due to fiber
saturation effects, while the faint limit corresponds to that of the
main galaxy sample ~\citep{str02}.

Following \citet{par05}, we define a ``Best'' volume-limited sample of
SDSS galaxies that contains the maximum number of galaxies, with
absolute-magnitude limits $-21.6 < M_r < -20.2$ that correspond to a
redshift range of $0.05 < z < 0.12$, or a comoving distance range of $
148.5 < r < 350.1~ h^{-1}$ Mpc, when the apparent magnitude cuts are
applied.  The definition of our ``Best'' volume is shown in Figure
~\ref{vol} as a dashed line.  The sky area covered by the parent
spectroscopic sample is illustrated in Figure 1 of Abazajian et
al. (2004).  It is important to note that the subsets of
spectroscopically-defined AGN drawn from this volume-limited galaxy
sample will not themselves be truly volume-limited samples because of
selection effects induced by their spectroscopic definition; the host
galaxies are restricted to the described range of absolute magnitude
and redshift, but the redshift distributions of AGN may not be uniform
because of the requirement that the emission lines be detected.  These
effects are discussed in detail in Section ~\ref{prop} and are taken
into account in our clustering analysis.

The spectroscopic subsamples of interest for this study suffer from
the fiber collision problem. When two targets lie closer than
55$\arcsec$, only one member of such galaxy pair will have a spectrum
available, unless later spectroscopic tiles overlap that same area of
sky.  Because classification as an AGN requires a spectrum for each
object, we cannot make any ``correction'' for such collisions, as is
commonly done for large-scale structure studies of SDSS galaxy
samples \citep{zeh02}.  The impact of this problem on our estimates of
the correlation functions is discussed later in Section ~\ref{csi}.

\subsection{Subsample Definition} 

\subsubsection{Active vs. Non-Active: the Spectral Classification} \label{samples}


The fraction of galaxies with an AGN is difficult to assess
observationally, because it strongly depends on the quality and
quantity of the available spectral data. Sub-classifying emission-line
nebulae typically requires a comparison of pairs of line-flux ratios
of spectral features that are relatively insensitive to reddening and
metallicity, are able to distinguish between different ionization
levels, and provide a useful segregation of different object types.
There is a set of line ratios that satisfy these criteria, and that
have been widely used in quantifying emission properties and
distinguishing between accretion-powered and starlight-powered
systems: [\ion{O}{3}]$\lambda$5007/H$\beta$,
[\ion{N}{2}]$\lambda$6583/H$\alpha$,
[\ion{S}{2}]$\lambda\lambda$6716,6731/H$\alpha$, and
[\ion{O}{1}]$\lambda$6300/H$\alpha$ \citep{bpt81, vo87}.

\begin{figure}
\plotone{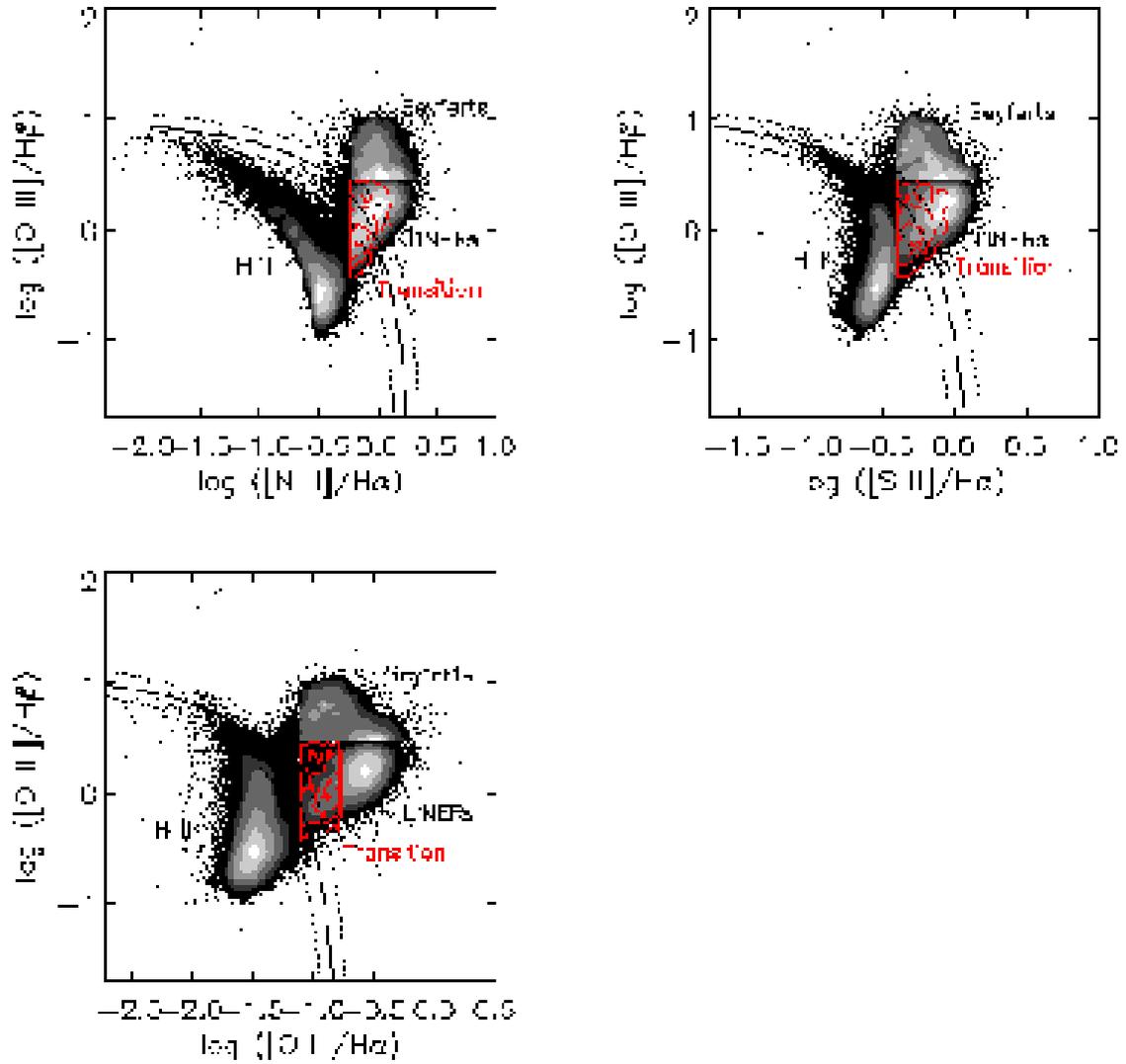}
\caption{ Emission-line diagnostic diagrams with separate density
contours for the four subclasses of narrow emission line galaxies that
have $all$ 6 features (H$\alpha$, H$\beta$, [\ion{O}{3}],
[\ion{N}{2}], [\ion{S}{2}], and [\ion{O}{1}]) detected in emission at
2$\sigma$ significance.  The transition objects are shown in red.
Density contour lines correspond to factors of $n$ of the total number
of objects in each class, where $n = 0.1, 0.2, 0.3, 0.5, 0.7, 0.9$
starting from outermost contour.  The contours corresponding to the
Transition objects are overplotted in red, and show the lines with $n
= 0.1, 0.3, 0.6, 0.9$.  The solid curve is the \citet{kew01}
theoretical separation between accretion and star-forming systems, and
the dotted lines are drawn at $\pm 0.1$ dex of this prediction.
\label{bpt}}
\end{figure}

Figure ~\ref{bpt} illustrates how galaxies that exhibit all 6 of these
emission lines at moderately high signal-to-noise, e.g., with
fractional errors $\leq$50\%, segregate as a function of these
combinations of line ratios.  Following the \citet{ho97} criteria, as
listed in Table~\ref{tbl-0}, we classify the objects as H {\sc ii}
galaxies, Seyferts, and LINERs, with their Pure LINER and Transition
subsets.  H {\sc ii}'s are the galaxies that populate mainly the left
branches of these diagrams, as they usually show weak low-ionization
transitions in [\ion{N}{2}], [\ion{S}{2}], and especially
[\ion{O}{1}].  By contrast, Seyferts and LINERs occupy the right
branch as they exhibit relatively strong forbidden features.  The
distinction between Seyferts and LINERs is usually given by their
ionization level, which is easily gauged by the [\ion{O}{3}]/H$\beta$
ratio, with Seyferts showing larger values.  The Transition systems
are differentiated from the Pure LINERs based on their low
[\ion{O}{1}]/H$\alpha$ ratio.  Such divisions offer a basis for
categorizing the line-emitters. However, as is apparent from these
three heavily populated diagrams, the nebular properties span a
continuous range; no breaks appear in the diagrams to indicate a
clear-cut transition between one type to another.


There are also relatively successful theoretical attempts to separate
the accretion-type galaxy nuclei from star-forming sources, using
criteria that can be described using these same spectra diagnostics
diagrams. Based on models that account for large realistic
metallicity, ionization parameter, and dust depletion ranges,
\citet{kew01} find that simple hyperbolic curves can be used
to split the line-emitting galaxies into AGN and star-forming
galaxies. As can be seen from Figure ~\ref{bpt}, the curves
claimed as upper limits on the location of H {\sc ii}-like objects
exclude only a small fraction of LINERs from the AGN class, and are
otherwise reasonably consistent with the ~\citet{ho97} classification
criteria.


Previous studies of low luminosity narrow-lined AGN in SDSS samples
used simpler criteria than we use here.  Based on the shape of the
distribution of objects in the [\ion{N}{2}]/H$\alpha$
vs. [\ion{O}{3}]/H$\beta$ diagram, \citet{kau03} propose an empirical
cut between AGN and H {\sc ii} -type galaxies, while some other
studies ~\citep{mil03, wak04} consider AGN classification schemes
based only on values of the [\ion{N}{2}]/H$\alpha$ flux ratio (i.e.,
with the condition that [\ion{N}{2}]/H$\alpha$ $> 0.6$).  These
classifications overlook the emission-line behavior as a function of
[\ion{S}{2}]/H$\alpha$, and [\ion{O}{1}]/H$\alpha$, which, e.g.,
\citet{vo87} and \citet{ho97} argue that discriminate better than
[\ion{N}{2}]/H$\alpha$ between the young stars and AGN as dominant
ionization sources.  Of the three low-ionization lines, [\ion{O}{1}]
is the most sensitive to the shape of the ionizing spectrum, requiring
a significantly hard radiation field, i.e., that of an AGN, to sustain
a sufficiently extensive partially ionized zone in clouds optically
thick to Lyman continuum, and thus to produce a strong such feature.
Since the ionization potential of [\ion{O}{1}] matches that of H very
well, large differences in the [\ion{O}{1}]/H$\alpha$ ratios are
expected between H {\sc ii} region-like objects and narrow line AGN.
The effect is also important for [\ion{S}{2}]
($\lambda$6717+$\lambda$6731)/H$\alpha$, however, the fact that S$^+$
can also exist within H$^+$ zones of H {\sc ii} regions and AGN
attenuates the difference between the two classes of objects.  Strong
[\ion{S}{2}] $\lambda\lambda$6717, 6731 are nevertheless expected in
AGN because of the large collisional-excitation cross sections for
these lines.  [\ion{N}{2}] has a much smaller excitation cross
section, and thus its strength in AGN is weaker.  Large
[\ion{N}{2}]/H$\alpha$ line ratios in AGN may however also be caused
by potential selective enhancements of nitrogen ~\citep{sto89, sto90},
and thus do not necessarily arise through ionization by an
accretion-type radiation field.

\begin{figure}
\plottwo{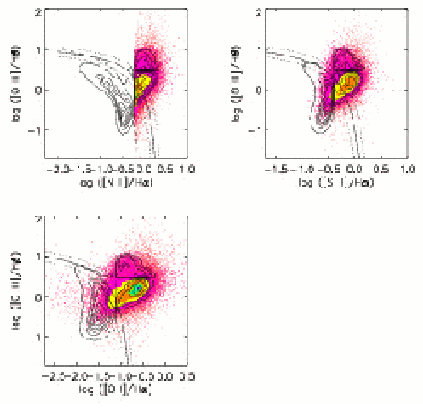}{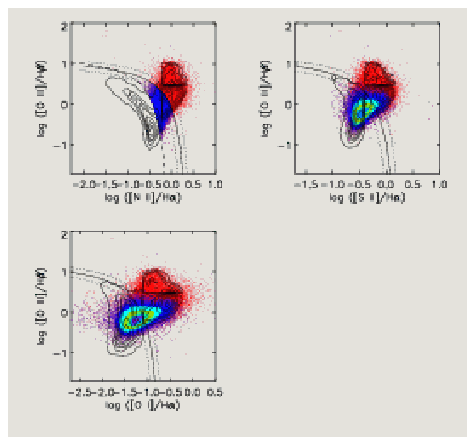}
\caption{ Emission-line diagnostic diagrams showing how the AGN
  defined using definitions of Miller et al. ({\it left panels},
  ``2-line'' AGN in red, and ``4-line'' AGN in magenta and
  multicolored density contours), and Kauffmann et al. ({\it right
  panels}, red $+$ blue points) distribute as a function of
  [\ion{S}{2}]/H$\alpha$ and [\ion{O}{1}]/H$\alpha$ line-flux ratios.
  For the latter case, the objects situated below the Kewley et
  al. curve in the [\ion{N}{2}]/H$\alpha$ vs. [\ion{O}{3}]/H$\beta$
  diagram, which are most likely to include contamination from
  star-forming activity, are overplotted in blue in this diagram, and
  are shown in multicolored contours in the other two diagrams.  The
  density contours presented in Figure ~\ref{bpt} are shown in the
  background for comparison.
\label{bpt2}}
\end{figure}

Figure ~\ref{bpt2} illustrates the behavior as a function of
[\ion{S}{2}]/H$\alpha$ and [\ion{O}{1}]/H$\alpha$ of the objects
classified as AGN based on [\ion{N}{2}]/H$\alpha$ ratio like in Miller
et al. (right panels), and based on the Kauffmann et al. line that
separates the two branches of the objects distribution in
[\ion{N}{2}]/H$\alpha$ vs. [\ion{O}{3}]/H$\beta$ diagram (left
panels).  For the Miller et al. classification, the density contours
show that these AGN are governed by a LINER-like behavior. Using the
spectral types defined in Table ~\ref{tbl-0}, the fraction of LINERs
is at least 30\%, while Seyferts comprise less than 9\% of these AGN,
in both their `2-line'' AGN (only detection of 2 line-emissions,
[\ion{N}{2}] and H$\alpha$ is required) and ``4-line'' AGN (additional
detection of [\ion{O}{3}] and H$\beta$ emission lines is required)
samples.  Also, the fraction of Miller et al. AGN with low
[\ion{S}{2}]/H$\alpha$ and [\ion{O}{1}]/H$\alpha$ ratios, which are
more likely to be dominated by star-forming activity, is as high as
20\%, which represent more than double the number of Seyferts.  AGN
defined using the Kauffmann et al. criteria severely leak into the H
{\sc ii} locus in both [\ion{S}{2}]/H$\alpha$ and
[\ion{O}{1}]/H$\alpha$ diagrams. 27\% of these sources, or 45\% of the
objects situated between the Kewley et al. and Kauffmann et al. lines
(the blue dots in the [\ion{N}{2}]/H$\alpha$ vs. [\ion{O}{3}]/H$\beta$
diagram), would be classified as H {\sc ii} using the criteria in
Table ~\ref{tbl-0}.  Seyferts remain poorly represented in the
Kauffmann et al. sample as well (less than 5\%), while LINERs,
although constituting the majority of the potentially
accretion-dominated sources (25\%), are greatly outnumbered by H {\sc
ii}'s.

This comparison shows that the Kauffmann et al. and the Miller et
al. AGN classifications, although powerful through their great
statistics, may not best represent accretion dominated narrow-line
galaxies.  Such classifications include a significant fraction of
galaxies that show star-forming activity, and certainly, a mix of
different types of AGN.  For the purpose of understanding differences
between emission-line activity originating in star formation and black
hole accretion, a more stringent characterization of the dominant
power source, particularly as a function of [\ion{S}{2}] and/or
[\ion{O}{1}], is necessary.  

Note that in our classification of the emission-line systems we
weighted equally the three diagnostic diagrams, even if the error
distributions in the line ratios vary from diagram to diagram.  In
particular, the errors in the [\ion{O}{1}] fluxes and
[\ion{O}{1}]/H$\alpha$ flux ratios are generally larger than those of
the other stronger features.  However, for Seyferts in particular, the
[\ion{O}{1}] error distributions are significantly narrower than those
of other object types, suggesting that for this particular category
the classification is especially trustworthy.

\subsubsection{Our Samples} \label{ours}

We select from the parent sample of galaxies a subset of strong
``emission-line'' sources and a set of ``passive'' objects that show
insignificant, if any, line emission activity.  The emission-line
sources are galaxies that exhibit all six lines H$\alpha$, H$\beta$,
[\ion{O}{3}],[\ion{N}{2}] [\ion{S}{2}], and [\ion{O}{1}] in emission,
i.e. the line fluxes have positive values and the fluxes are measured
with at least 2$\sigma$ confidence.  We identify 54,410 such strong
line-emitters, which comprise $\approx 20$\% of all objects.  Note
that our definition of emission-line objects is very restrictive, and
therefore the detection fraction of such sources among all galaxies is
smaller than what \citet{ho97} and \citet{mil03} find ($\ga 40\%$).
When we relax the ``6-line'' definition criteria to include any kind
of emission-line activity, we recover a similarly high fraction.  The
``passive'' galaxies are defined here as the objects for which
H$\alpha$, H$\beta$, [\ion{O}{2}], and [\ion{N}{2}] are not detected
in emission (i.e., their EWs are positive in the emission-line
catalog, indicating an absorption feature). This sample, which
includes 59,197 sources, is a subset of the red, old, non-H$\alpha$
emitters.

Among the emission-line galaxies described above, we 
use the criteria listed in Table ~\ref{tbl-0} to identify 34,781 H {\sc
ii}-type galaxies, 2,392 Seyferts, and 7,468 LINERs, among which 2,155
are Transition objects.  There are a significant number of strong
emission-line objects ($\approx$18\%) that do not simultaneously
comply with these sets of conditions: these are objects that show
large [\ion{N}{2}]/H$\alpha$ ratios ($>$ 0.6) but low
[\ion{S}{2}]/H$\alpha$ ($<$ 0.4) and/or [\ion{O}{1}]/H$\alpha$ ($<$
0.08), or small [\ion{N}{2}]/H$\alpha$ and large ratios involving
[\ion{S}{2}] and/or [\ion{O}{1}].  For the sake of simplicity, we
exclude these sources from the comparative analysis of clustering of
Seyfert, LINER and H{\sc ii} populations.

Although [\ion{O}{1}] is generally weak and therefore difficult to
measure accurately, this feature is very useful in separating the
Transition objects among LINERs. Note, however, that requiring its
detection in the sample definition severely diminishes the sample
sizes. Without this condition on [\ion{O}{1}] and consequently on
[\ion{O}{1}]/H$\alpha$, we identify 98,556 line emitters, or $\approx
40$\% of the galaxies, among which 3,052 are Seyferts, 12,779 are
LINERs, and 65,971 are H {\sc ii} galaxies.  In our clustering
analysis, we compare the clustering properties of active galaxies
whose definitions both include and exclude conditions on [\ion{O}{1}]
and [\ion{O}{1}]/H$\alpha$.


\subsection{Subsample Properties \& Selection Effects} \label{prop}

\begin{figure}
\plottwo{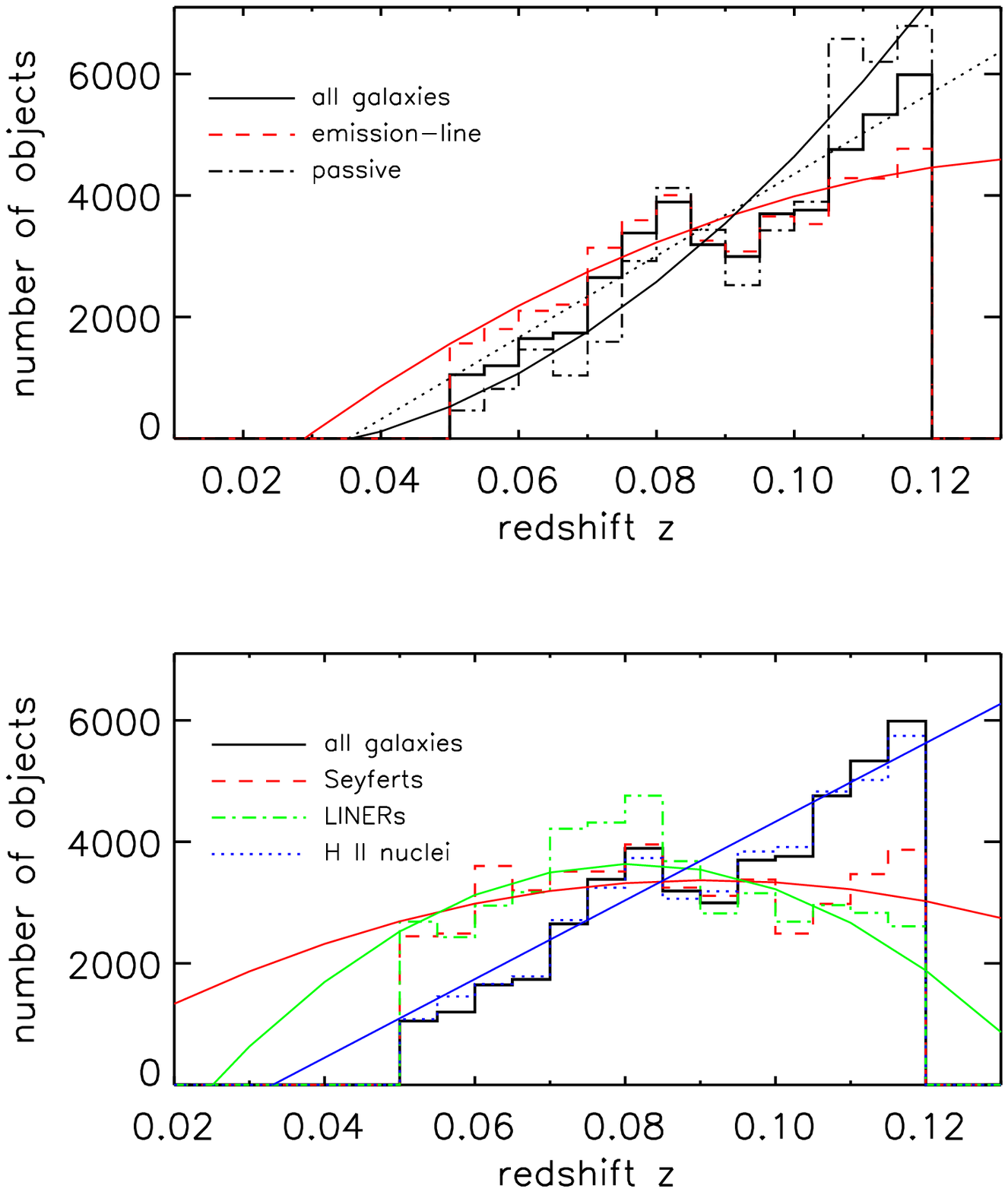}{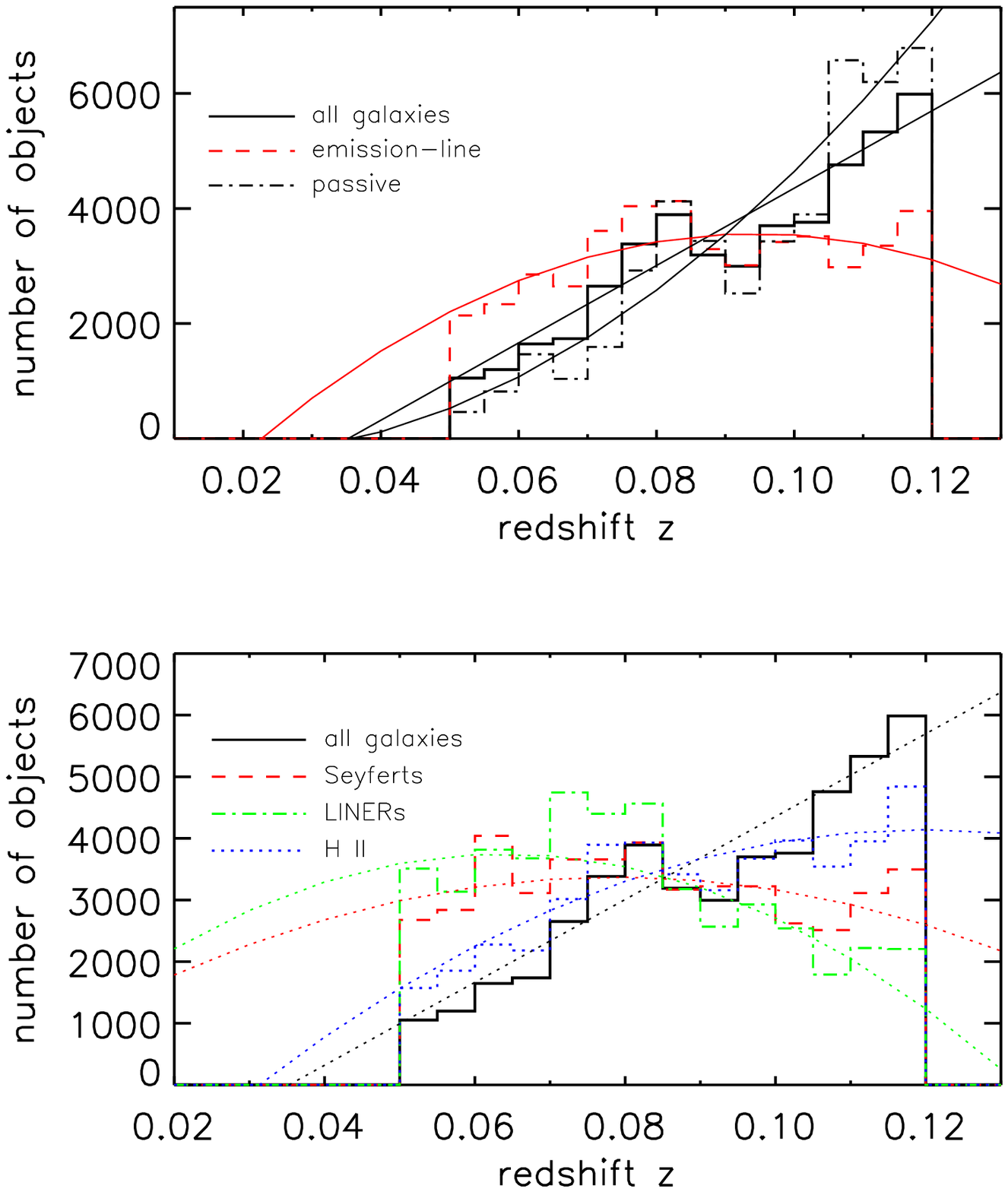}
\caption{ Redshift distributions for different subclasses of
  objects, in the absolute magnitude limited samples.  The {\it left}
  panels corresponds to spectral classification that does not include
  conditions on [\ion{O}{1}] and [\ion{O}{1}]/H$\alpha$ while the {\it
  right} panels show objects whose classification includes these
  parameters (note the lower number of the emission-line systems,
  especially at higher redshifts).  The histograms are re-normalized
  to have the total number of objects equal to that of the galaxy
  sample.  The solid curves show the corresponding 2nd order
  polynomial fits for subclasses of objects in the absolute magnitude
  limited samples, while the dotted line shows the polynomial fit for
  the volume limited galaxy sample.
\label{z}}
\end{figure}
\begin{figure}
\plottwo{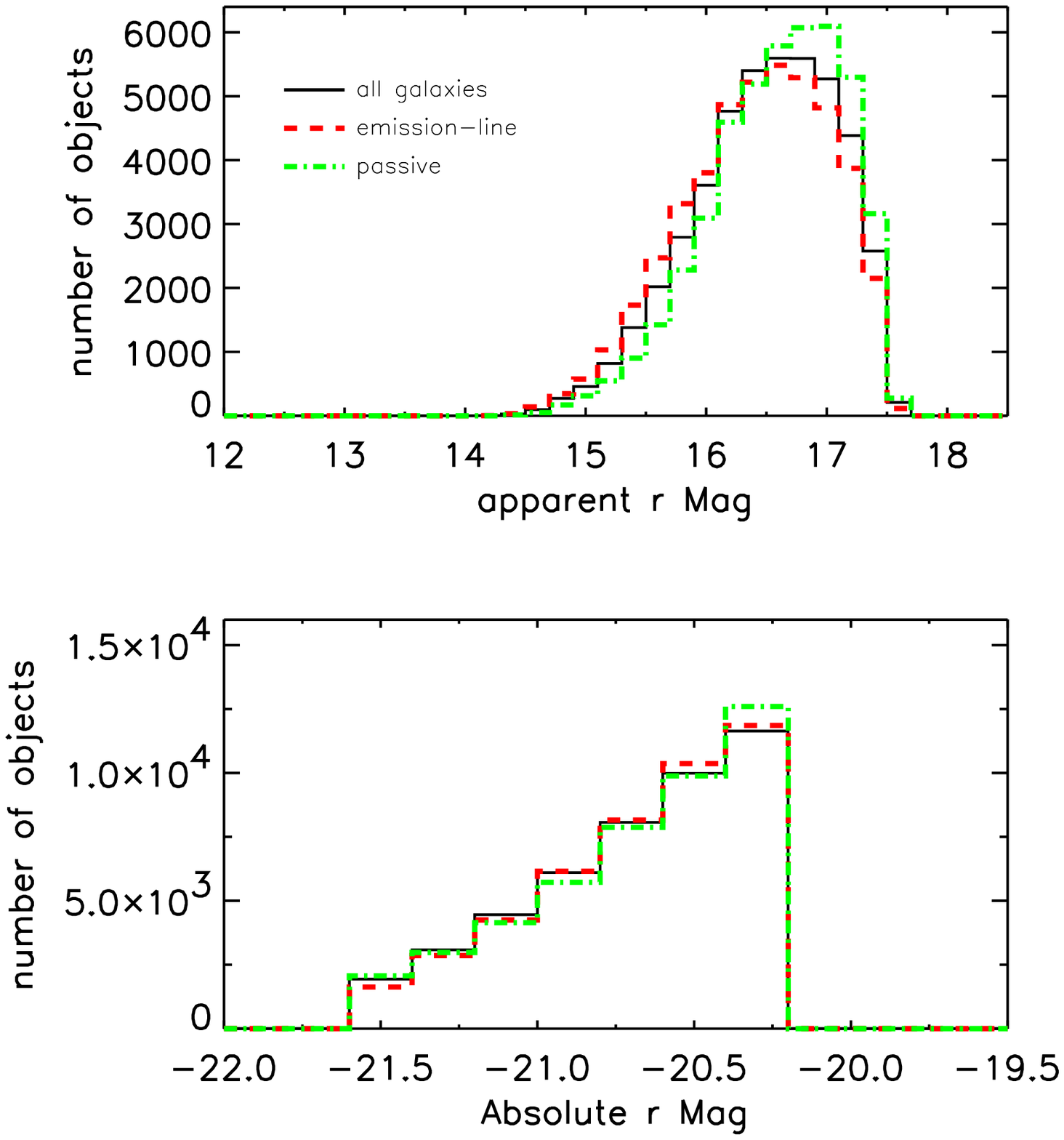}{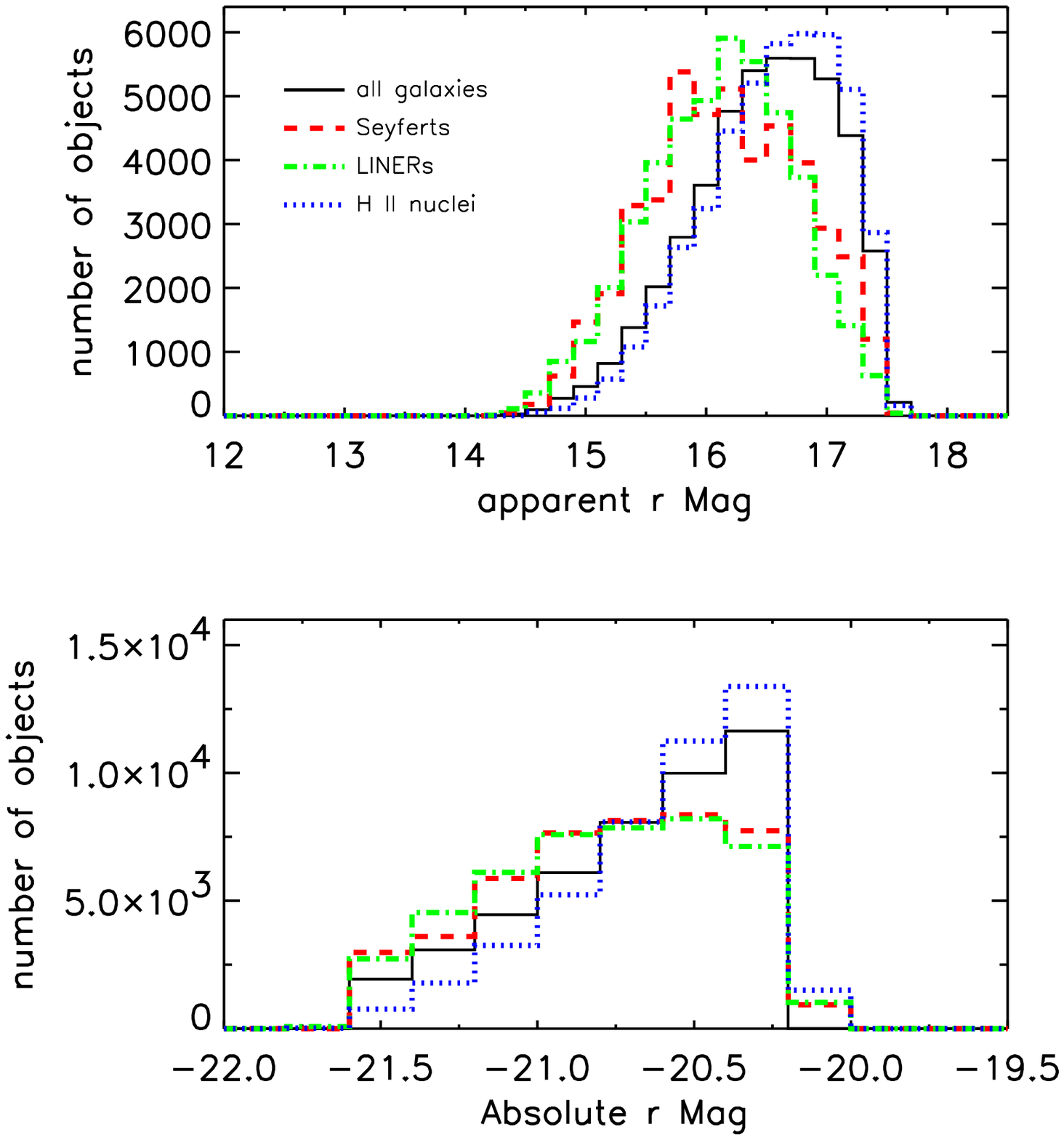}
\caption{ Distributions in apparent and absolute $r$ magnitude of the
  hosts of different spectroscopically-selected subclasses, all of
  which were drawn from a volume-limited parent galaxy sample.  The
  histograms are re-normalized to have the total number of objects
  equal to that of the galaxy sample.
\label{Mag}}
\end{figure}

We begin by forming the ``Best'' absolute-magnitude limited sample of
galaxies using the definition illustrated in Figure ~\ref{vol}, then
apply the criteria listed in Table ~\ref{tbl-0} to form
spectroscopically-classified subsamples. The numbers of objects in
each subsample are recorded in Table ~\ref{tbl-1}.  Figure ~\ref{z}
shows the redshift $z$ distributions for these samples and comparisons
of their apparent and absolute $r$ magnitudes are presented in Figure
~\ref{Mag}. 

It is clear that, although drawn from a parent sample of galaxies with
a fixed range of intrinsic luminosities, subsamples of
spectroscopically-classified objects are not, in general, volume
limited.  The necessity of detecting the spectral features employed in
classification causes a redshift dependence. For example, LINERs and
Seyferts are less prominent than other types of sources at $z > 0.1$,
and seem to populate the brighter galaxies, both in apparent and
absolute magnitudes, while the H {\sc ii} systems are shifted toward
fainter apparent and absolute magnitudes.

Whether a specific galaxy is included into a given spectral category
is strongly subject to multiple selection effects that vary with
redshift.  The fraction of galaxy centers exhibiting measurable
emission lines decreases with redshift as a consequence of decreasing
signal-to-noise. More importantly, the increase with distance of the
projected size of the SDSS 3$\arcsec$ spectroscopic aperture causes
dilution of emission-line activity by the increasing amount of host
galaxy light.  Thus, weak line-emitting systems are more likely to be
misclassified as non-active at larger distance.  These effects explain
the differences seen in the redshift and magnitude distributions of
the accretion type sources (Seyferts, and at least some LINERs) and
the galaxies in which activity is mainly driven by hot young stars (H
{\sc ii}).  While in the latter objects the power source is more
extended, and thus more likely to be observed at larger distances, in
the accretion dominated nuclei the ionization originates in more
centrally concentrated regions and thus is less likely to be included
in the emission-line sample at large $z$ where the signal-to-noise is
lower.


Note that for our specific goals of separating as cleanly as possible
among different types of emission-line systems, the fixed aperture
biases remain inconsequential.  Regarding the detection of the AGN
dominance, the fixed fiber effects go only in one direction: the more
distant objects are less likely to be detected as AGN as stellar
contamination becomes more pronounced; hence, such cases would most
likely account for the large fraction of objects that remain
unclassified based on our criteria, and which we exclude from our
comparative analysis between properties of bona-fide Seyferts and
LINERs.  Likewise, we do not attempt any aperture correction for the
line fluxes as estimates of total quantities (like e.g., the star
formation rate) remain irrelevant for this work.  Note also that the
relative number statistics of various types of emission-line systems
are indeed expected to be different for different $z$ bins for a fixed
fiber size.  However, such possible classification biases are
minimized through our choice of volume-absolute magnitude limited
sample that spans relatively short ranges in both redshift and
brightness.


We find that the different spectral classes of objects exhibit
distinct behavior in several physical characteristics. Figure
~\ref{o1o3} shows the distributions in log $L_{\rm [O I]}$ and log
$L_{\rm [O III]}$, the intrinsic column density $N_H$, and the
electron density $n_e$, separately for the Seyferts, LINERs, H {\sc
ii} galaxies, and the whole sample of 6-line emitting galaxies.  Line
luminosities $L_{\rm [O III]}$ and $L_{\rm [O I]}$ are calculated only
for objects that show flux measurement uncertainties of $< 50$\% in
these lines and in the Balmer decrement H$\alpha$/H$\beta$, which is
used to correct for intrinsic extinction.  We assume a dust-free
case-B recombination value of 2.86 for the standard, not absorbed
H$\alpha$/H$\beta$, and, for the sake of simplicity and to ease
comparison with previous calculations by \citet{kau03}, an attenuation
law of the form $\tau \propto \lambda^{-0.7}$ \citep{cha00}.  We use
the same method for calculating the $N_H$ values.  To estimate $n_e$
we employ the [\ion{S}{2}]$\lambda$6716/$\lambda$6731 line-flux ratio
\citep{ost89}.  The [\ion{O}{2}]$\lambda$$\lambda$3729,3726 feature
could also be used as a density indicator, as it is available in the
majority of the SDSS galaxy spectra, and the emission-line catalog
that we use here includes measurements of the individual components.
However, the small separation of only 3\AA\ between the two lines may
cause large uncertainties in their fluxes, and therefore in the $n_e$
estimates as well.

Investigation of the $L_{\rm [O III]}$-dependent behavior of different
types of galaxies is important because this parameter is claimed to
trace the AGN activity, at least for luminous systems with large
[\ion{N}{2}]/H$\alpha$ ratios \citep{kau03}.  This work showed
evidence for the fact that highly [\ion{O}{3}] luminous objects are
more likely to be powered by an accretion-type nucleus, suggesting
that larger $L_{\rm [O III]}$ may correspond to larger accretion
rates.
A scrutiny of the role played by $L_{\rm [O I]}$ in distinguishing
among different spectral types and in their spatial clustering
properties is motivated by the strong capability of the [\ion{O}{1}]
feature to differentiate between thermal and non-thermal ionization
sources.  Given that [O I] arises preferentially in zones of partly
ionized hydrogen, that are extended only in objects photoionized by a
spectrum containing a large fraction of high-energy photons, i.e.,
originating in accretion, and therefore nearly absent in galaxies
photoionized by OB stars, one might expect that $L_{\rm [O I]}$ is
also be a good indicator of the strength of the accretion activity.
Such expectations are supported by recent work of \citet{con05} who
indicate that the presence and strength of the [\ion{O}{1}] emission
may be linked to the presence and strength of the broad H$\alpha$
emission, and thus to the dominance of the accretion-type ionization
in the nuclear spectra of nearby emission-line galaxies.

Interestingly, in terms of both $L_{\rm [O III]}$ and $L_{\rm [O I]}$,
LINERs and Seyferts show a strongly antagonistic behavior.  LINERs
stand out as the objects with the lowest $L_{\rm [O III]}$ and $L_{\rm
[O I]}$ relative to other spectral classes, rarely exhibiting values
of log $L_{\rm [O III]}$ $> 40$, with $L_{\rm [O III]}$ in units of
erg s$^{-1}$.  Seyferts, on the other hand, are the galaxies that are
most likely to be luminous in these emission lines; the median values
for log $L_{\rm [O III]}$ are $\sim$38.6 for LINERs, and $\sim$39.8
for Seyferts.  Although the lack of LINERs with high $L_{\rm [O III]}$
can be explained by their definition as low ionization emission
regions, which explicitly selects lower [\ion{O}{3}] fluxes, the
absence of high $L_{\rm [O I]}$ LINERs is intriguing. Because the
redshift and magnitude distributions of Seyferts and LINERs are very
similar, selection effects cannot be attributed to this trend, thus
their low $L_{\rm [O III]}$ and low $L_{\rm [O I]}$ must be intrinsic
to these systems.

Dust obscuration is another variable that could significantly affect
the detection of AGN-like activity.  In terms of intrinsic
attenuation, as measured by $N_H$, LINERs are again at one extreme,
showing low levels of extinction relative to Seyferts and H {\sc
ii}'s, with the latter being the most obscured of the emission-line
objects.  These trends are to be expected, as star formation is
generally accompanied by significant amount of obscuring
dust. However, it may also be the case that in such a dusty
star-forming environment, a potentially present weak AGN-like nuclear
emission is suppressed, because it lies buried in surrounding H {\sc
ii}- like features.

Based on the [\ion{S}{2}] line flux ratio estimates for the electron
density, Seyferts show the highest $n_e$, as expected for sources in
which the dominant ionization mechanism is accretion of matter onto a
black hole.  The LINERs show lower $n_e$ densities, while the
starforming systems clearly populate the lowest end in the $n_e$
distribution; the median $n_e$ is $\approx 5 \times 10^2$ cm$^{-3}$ ,
$\approx 1.5 \times 10^2$ cm$^{-3}$ , and $\approx 5 \times 10^1$
cm$^{-3}$ for the Seyferts, LINERs and H {\sc ii}'s respectively.
A Kolmogorov-Smirnov test indicates an almost null probability that
any pair of these subsamples draw their distributions in [\ion{S}{2}]
ratio from the same parent distribution; the KS probability that
Seyferts and the whole sample of emission-line emitters belong to the
same parent population is KS$_{\rm Sey/emiss} = 4.24 \times 10^{-36}$,
while for the comparison between the line emitters and the H {\sc
ii}'s, KS$_{\rm emiss/H II} = 2.42 \times 10^{-23}$.  There are
roughly 3000 galaxies that manifest emission-line activity with
$n_e < 10$ cm$^{-3}$, which is below the theoretical lower limit for
which $n_e$ can be estimated based on [\ion{S}{2}] emission; about
half of these objects are H {\sc ii} galaxies.

Median and average values of the parameters for which the
distributions are compared among different spectral types of objects
are recorded in Table ~\ref{tbl-param}.

\section{The Correlation Function} \label{csi}

To quantify clustering in the observed galaxy distribution we estimate
the 2-point correlation function (CF) $\xi(s)$, which measures the
excess probability over random to find an object pair separated by
comoving distance $s$. The separation $s$ is measured in
redshift-space, thus it is important to note that the shape and
amplitude of this measure of spatial clustering are determined not only
by the spatial configuration but also by peculiar velocities. Small
scale velocity dispersion, e.g., in cores of galaxy clusters, damps
small scale clustering, while linear infall onto dense regions
amplifies clustering on large scales.

Accurate estimation of the correlation function requires careful
accounting for both angular and radial variation in the expected
galaxy density.  The former requires knowledge of the complicated
geometry of the angular mask that describes the regions on the sky in
which data were obtained and the sampling of galaxies in the observed
regions, while the latter variation may depend on physical
characteristics of the samples analysed, which can introduce important
selection biases with distance.



To account for the survey geometry we employ random catalogs that
describe the SDSS DR2 survey angular selection function. These random
masks are generously made publicly available in the NYU Value-Added
Galaxy Catalog and described in ~\citet{bla04}.  The random samples we
employ in all the $\xi(s)$ measurements are at least 10 times as large
as the datasets involved.  As discussed in Section ~\ref{samples}, the
$z$ distributions for the volume limited samples of the different
categories of objects are not uniform, and are different from sample
to sample.  We therefore build individual radial selection functions
for the random masks corresponding to each sample to be studied; we
randomly assign redshifts that follow distributions constructed based
on 2nd order polynomial fits to the observed object $z$ distributions.
We have explicitly tested the CF $\xi(s)$ estimates for different
parametrizations of the redshift distributions $N(z)$, and find that
they do not significantly affect the results; the power-law fits,
described below, show slopes and amplitudes that remain consistent
within the value range bordered by uncertainties.


As discussed above, we do not account for the potential biases on
small scales introduced by the fiber collisions, because we have no
reasonable means of doing so.  In other estimates of galaxy clustering
properties, such biases have been corrected for by assigning each pair
member whose redshift was not obtained because of the fiber collision
the same $z$ as the pair member whose redshift was measured. However,
for the purpose of studying the dependence of clustering on spectral
characteristics we cannot apply such correction because there is no
physical motivation for assigning to the unobserved pair member the
spectral properties found to characterize the object whose spectrum is
available.  Therefore we treat the fiber collisions as a contribution
to incompletness in the survey; this variation of the completeness
with angular position of the is accounted for in the random catalogs
described above.  This procedure leaves a residual effect due to fiber
collisions that is significant at very small scale, i.e., at comoving
transverse separations of $\sim$0.14$h^{-1}$Mpc (the corresponding
linear scale for 55$\arcsec$, at the outer edge of our samples $cz$ =
50,100 kms$^{-1}$).  We restrict our measurements to separations $>
0.14h^{-1}$Mpc and note that the incompletness biases due to fiber
collisions should similarly affect all of the subsamples we analyse,
and thus are not a significant concern for this study because our
focus is on the relative differences between the CF characteristics.


We calculate the correlation functions using the \citet{lan93}
estimator, 
\begin{equation}
\xi(s) = 1 + \Big(\frac{N_{rd}}{N}\Big)^2 \frac{DD(s)}{RR(s)} -2
\frac{N_{rd}}{N} \frac{DR(s)}{RR(s)},
\end{equation}
using bins in pair separation $s$ that are logarithmically spaced with
a width of 0.33 in log($s/h^{-1}$Mpc), starting from 0.147$h^{-1}$Mpc.
In this formula, $DD(s)$ and $RR(s)$ are the number of pairs in the
data and random catalogs respectively, while $DR(s)$ is the number of
pairs between the data and the random samples.  Pair counts are
estimated using the computationally efficient ``npt'' algorithm
described in \citet{npt}.  We estimate the covariance matrices of the
uncertainties using the statistical jackknife method \citep{lup93},
for which we divide our samples into 75 separate regions on the sky,
of approximately equal area, $\backsimeq 10\deg \times 5\deg$, and
estimate the CF 75 times, each time leaving out one of these regions
(see \citet{zeh02} for a more detailed description of this method and
for a discussion on its robustness in estimating the uncertainties).


We quantify the clustering properties of a given sample by
fitting the measured CF with a power law of the form
\begin{equation}
\xi(s) = \big(s/s_0\big)^{-\gamma},
\end{equation}
where $s$ denotes the comoving separation between the object pairs,
$s_0$ is the correlation length used to express the amplitude of
clustering, and the slope $\gamma$ quantifies the ratio of
small to large scale clustering.  In the fitting process we employ
separations $3 h^{-1}$ Mpc $< s < 10 h^{-1}$ Mpc, in order to avoid
separations on which the CFs are derived from fewer than 20 data-data
pairs.

The CF of the full volume limited ``Best'' galaxy sample is
characterized by a correlation amplitude $s_0 = 7.8 h^{-1}$ Mpc, and a
slope $\gamma \approx 1.2$.  These values are consistent with previous
calculations of the redshift-space clustering of the SDSS main galaxy
sample, both for magnitude and volume limited samples similarly
bounded in luminosity \citep{zeh02, zeh05}.

\subsection{Clustering Dependence on Spectral Properties} 

We examine here the redshift-space CF for various subsamples
constructed based on sets of properties that reflect differences in
the dominant ionization mechanism, i.e., accretion vs. star-forming
type, or provide clues to the physical conditions of the line-emitting
regions. 
We compare correlation functions of different spectral
classes, and investigate the degree to which other emission-line
related parameters, e.g., [\ion{O}{3}] and [\ion{O}{1}] luminosities,
intrinsic neutral hydrogen column densities ($N_H$), and electron
densities $n_e$, may influence the results.

The results of power-law fits to the CFs estimated for various
subsamples of the parent ``Best'' absolute magnitude limited galaxy
sample considered in this analysis are recorded in Table ~\ref{tbl-1},
together with results for the CF of all galaxies in the sample.
Object sample definitions as a function of $n_e$ and $N_H$ are
described in Tables ~\ref{tbl-ne}, and ~\ref{tbl-nh}.  It is clear
that the clustering properties of subsamples of objects governed by
distinct spectroscopic characteristics are different. In the following
subsections we discuss these results in detail.

\subsubsection{Spectral Class} \label{class}

The CFs estimated for emission-line objects, passive objects, and the
three distinct line-emitting subsets of Seyferts, LINERs and H {\sc
ii}'s, are illustrated in the upper and lower panels of Figure
~\ref{tpcf_class}. We present and compare the shape and amplitude of
power-law fits to the CFs for samples whose definition both include
and exclude constraints on [\ion{O}{1}] and [\ion{O}{1}]/H$\alpha$, in
the right and left panels respectively.  The CF of the volume limited
sample of all galaxies is also shown for comparison in all panels.
The error bars denote the 1-sigma uncertainties.  The upper panels
include insets that compare the 1-sigma, 90\%, and 99\% confidence
contour levels for the likelihood functions obtained by fitting
power-law models to the correlation functions of actively
line-emitting galaxies and the passive ones.

The passive galaxies have a substantially higher amplitude and steeper
$\xi(s)$ than the emission-line galaxies, with a correlation length
$s_0 \approx 11.4 h^{-1}$ Mpc, compared to $s_0 \approx 6.5 h^{-1}$
Mpc for the actively line-emitting sources.  This trend is clearly
present whether or not the selection of these objects employ the
[\ion{O}{1}] feature, although the datasets are smaller when
[\ion{O}{1}] is required in defining the sample and, therefore, the
uncertainties are correspondingly larger.

Among the emission-line sources, LINERs exhibit the highest clustering
amplitude, while H {\sc ii}'s show the weakest clustering, at both
small and large scales (see lower panels of Figure ~\ref{tpcf_class}).
Seyferts distinguish themselves from the other emission-line galaxies
by the relatively steep slope of their correlation function, $\gamma >
1.4$.  This effect might arise if the environment of Seyferts
causes an excess (relative to a flatter power law) of AGN activity on
a scale larger than clusters or groups; the power-law fits are over
separations $3-10h^{-1}$Mpc, so the excess small-scale clustering is
not caused by very close pairs of Seyferts.  Overall, Seyferts'
clustering amplitude remains intermediate between that of LINERs and H
{\sc ii}'s.

A more quantitative comparison between the clustering properties of
Seyferts, LINERs, H {\sc ii}'s, and of the whole galaxy sample is
presented in Figure ~\ref{csi_class}.  From top to bottom, we see that
the clustering amplitude of Seyferts is significantly smaller than
that of the full galaxy sample, LINERS are clustered similarly to the
parent galaxy sample, and HII's are less clustered than all
galaxies. The left and right sets of panels are for AGN sub-samples
defined without employing [OI] and with [OI], respectively. Because
this criterion changes the number of objects in the sub-samples, as
well as changing the intrinsic luminosity necessary for this line to
be detected at fixed distance, we find slight variation in the
statistical significance of these comparisons.

That Seyferts are found to be less clustered than the average galaxies
is an important result, and its statistical significance requires some
comment.  Table ~\ref{tbl-1} shows that the clustering amplitudes for
Seyferts and for all galaxies in the main sample are $s_0 = 6.00 \pm
0.64$ and $7.80 \pm 0.49$ respectively, which translates into a
difference of order 3$\sigma$.  In the upper panel of Figure
~\ref{csi_class} one sees that the 2-dimensional "1-$\sigma$" limits
just touch, and the 90\% contours overlap.  This overlap arises
because the 2-dimensional contours enclose the labeled percentage of
the probability in the joint distribution of both $s_0$ and $\gamma$,
and these elliptical contours extend beyond the confidence limits that
are computed in one parameter at a time (by marginalizing over the
other parameter).  The natural confidence regions in the 1-dimensional
parameter space, say $s_0$, are the {\it projections} of the
2-dimensional regions defined by fixed $\Delta \chi^2$ into this
1-dimensional space of interest.  To enclose the same probability,
i.e., fraction of points, say 68.3\% (or $\Delta \chi^2 = 1$), as the
two parameters $s_0$ and $\gamma$, the corresponding ellipse will
correspond to $\Delta \chi^2 = 2.3$ as it must necessarily extend
outside of both of them \footnote{e.g., see discussion in Chapter 15
of Press et al. 1992, ``{\sl Numerical Recipes in C},''
http://www.library.cornell.edu/nr/bookcpdf.html}.  The resulting
constraints on $s_0$ and $\gamma$ are reported in Table ~\ref{tbl-1}.

To clarify the statistical significance of the observed difference in
clustering amplitude of Seyferts and the parent sample of galaxies, we
conduct an additional test.  Because the Seyferts are a sub-sample of
the volume-limited sample of galaxies, a simple test of the validity
of our results is to randomly select sub-samples of galaxies with the
same number of objects as the Seyfert dataset, compute the best-fit CF
parameters for each, and compare the results with the actual Seyfert
CF.  We form 100 such ``mock Seyfert'' samples for two separate cases:
1) by randomly selecting 1000 objects from the parent galaxy
population, and 2) additionally constraining the mock Seyfert sample
to have the same redshift distribution as the Seyferts.  For each mock
sub-sample we fit the correlation functions using the covariance
matrix built from the variation among the 100 CFs. The results shown
in Figure ~\ref{mock} indicate that the CF of the Seyfert sample is
inconsistent at $>95\%$ confidence with being drawn from a parent
population of CF's of randomly selected galaxies. In both cases, not a
single random sample of galaxies has a clustering amplitude $s_0$ as
low as observed in the actual Seyfert sample.

It is interesting to note that, while for the samples of Seyferts and
LINERs that use [\ion{O}{1}] in their definitions the error bars are
larger than in the case when this feature is not used, the differences
in $s_0$ and $\gamma$ become more pronounced, and more statistically
significant: LINERs' $s_0$ is in this case indistinguishable from that
of the whole sample of galaxies, while Seyferts' $s_0$ is much lower
and the 1-$\sigma$ contours do not touch anymore.  On the other hand,
$s_0$ of the star-forming galaxies does not actually change between
samples that use [\ion{O}{1}] or not in their definition.  This is yet
another clue to the fact that the [\ion{O}{1}] emission is not an
important characteristic of the H {\sc ii} galaxies, while it
certainly makes a difference in defining more strictly AGN, i.e.,
Seyferts and LINERs.

Interestingly, regarding the CF characteristics of the data samples of
different spectral types we find a slight tendency toward steeper
slopes, as compared to that of the galaxies' CF.  Note also that the
correlation functions of the sub-samples do not necessarily average to
the correlation function of the whole sample; the CF of the whole
sample is a weighted sum over the correlations of the sub-samples and
the cross-correlations between the sub-samples.  It could be the case
that the CFs of the sub-samples are steeper because certain
environments favor the formation of particular species of AGN.  In an
extreme example, if Seyferts, LINERs, and H {\sc ii} galaxies came in
clumps of one type only, they would be strongly clustered on small
scales, still clustered on larger scales, but have a relatively
steeper correlation function because of the extra small-scale
clustering.  In contrast, the cross-correlations between sub-samples,
e.g. between Seyferts and H {\sc ii} galaxies, would have very little
small scale clustering because they would be mutually exclusive on the
scale of the clumps, thus the cross-correlation functions would be
shallower than that of the whole sample.

In order to examine our results in the context of previous work on
clustering of AGN observed by the SDSS, we also examine the CF of the
narrow-line AGN vs. non-AGN (i.e., star-forming sources) using the
classifications employed by Kauffmann et al., and Miller et al.
Figure ~\ref{agnKM} shows the CFs estimated for such subsamples of the
``Best'' volume limited galaxy sample and comparisons of the
confidence contours of power-law fits to these CFs.  The galaxy CF is
also shown for comparison.  The fitted power-law parameters and number
of objects in each sample are also listed in Table ~\ref{tbl-1}.  Here
we find that AGN samples selected using the criteria applied by
Kaufmann et al. and Miller et al. show very similar clustering
amplitudes to that of all galaxies, as claimed in previous studies
based on estimates of population fractions in cores of galaxy clusters
and intra-cluster fields \citep{mil03}, or measurements of CFs
\citep{wak04}.  This result is not surprising.  We show in Section
~\ref{samples} that AGN samples selected using these criteria are
dominated by LINERs. We also find that LINERs, as a
spectroscopically-distinct population, are indeed clustered similarly
to normal galaxies.  Thus, the resemblance between the spatial
clustering properties of such AGN samples and normal galaxies reflects
dominance of LINERs in these AGN definitions, and it does not apply to
AGN in general, particularly not to Seyferts as a separate class.
Similar trends were reported before by \citet{kau04} who also note
that the environmental effects in the Miller et al. sample are likely
to be dominated by LINERs.  We show here with great confidence that
the environments and the large scale structure of weak and powerful
AGN are distinct, and that if these systems are not carefully
separated out, one can erroneously characterize the AGN
population when instead refers to LINER-like objects only.

The results presented in this section indicate clearly that bona fide
narrow-line low luminosity AGN, i.e., the Seyfert 2s, are less
spatially clustered than typical galaxies.  These findings have
important implications for the environment of the host galaxies of
AGN.  The statistics of peaks in a Gaussian random field suggest (and
results of fully non-linear N-body simulations demonstrate) that a
large clustering amplitude is indicative of a population of objects
that inhabit high peaks (dense regions) and, conversely, that a low
clustering amplitude implies a sample that probes lower peaks of the
field and less dense environments.  Thus, we infer that Seyferts and H
{\sc ii} galaxies tend to prefer less crowded environments than
typical galaxies, while LINERs seem to follow the galaxy spatial
distribution, as their correlation function is, within the
uncertainties, identical to that of the whole galaxy sample.  The
cause of the relatively steeper slope of the CF of Seyferts is
unclear, but it may be further evidence of a special environment for
promoting their activity.  Our analyses seem to confirm the trends
reported by \citet{mil03}, which show a the predilection of the
star-forming galaxies to reside in more sparsely populated regions,
and of the passive, non-line-emitting galaxies to live in more dense
environments.

\subsubsection{[\ion{O}{3}], [\ion{O}{1}] line luminosities} \label{csio1o3}

Figure ~\ref{tpcf_o1o3} shows the CF of subsamples selected by
luminosities of the forbidden narrow emission lines [\ion{O}{3}] and
[\ion{O}{1}].  We refer to objects that have log $L_{\rm [O III]} >
40$ and log $L_{\rm [O III]} < 37$ as the high and low $L_{\rm [O
III]}$ samples respectively. Line luminosities are in units of erg
s$^{-1}$.  Conforming with the definitions listed in Table
~\ref{tbl-1}, a similar sample name designation is used for the low
and high $L_{\rm [O I]}$ object samples.  Our calculations show that
objects that are bright in [\ion{O}{3}] and/or [\ion{O}{1}] are
significantly less clustered than sources that are less luminous in
these lines.  We confirm the dependence of the clustering amplitude on
$L_{\rm [O III]}$, as seen by \citet{wak04}.  For the first time, we
detect an even stronger variation of clustering with $L_{\rm [O I]}$.

\begin{figure}
\plottwo{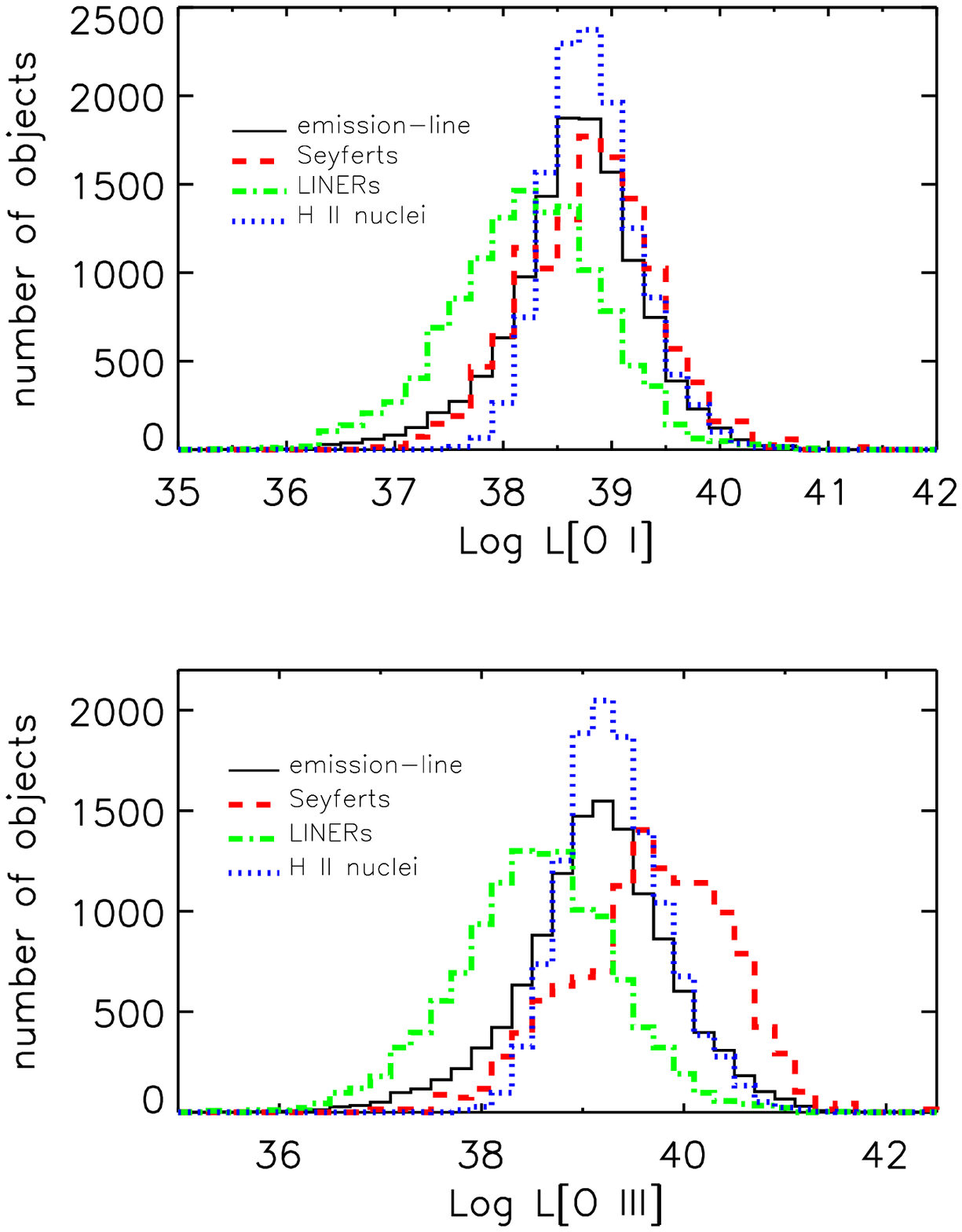}{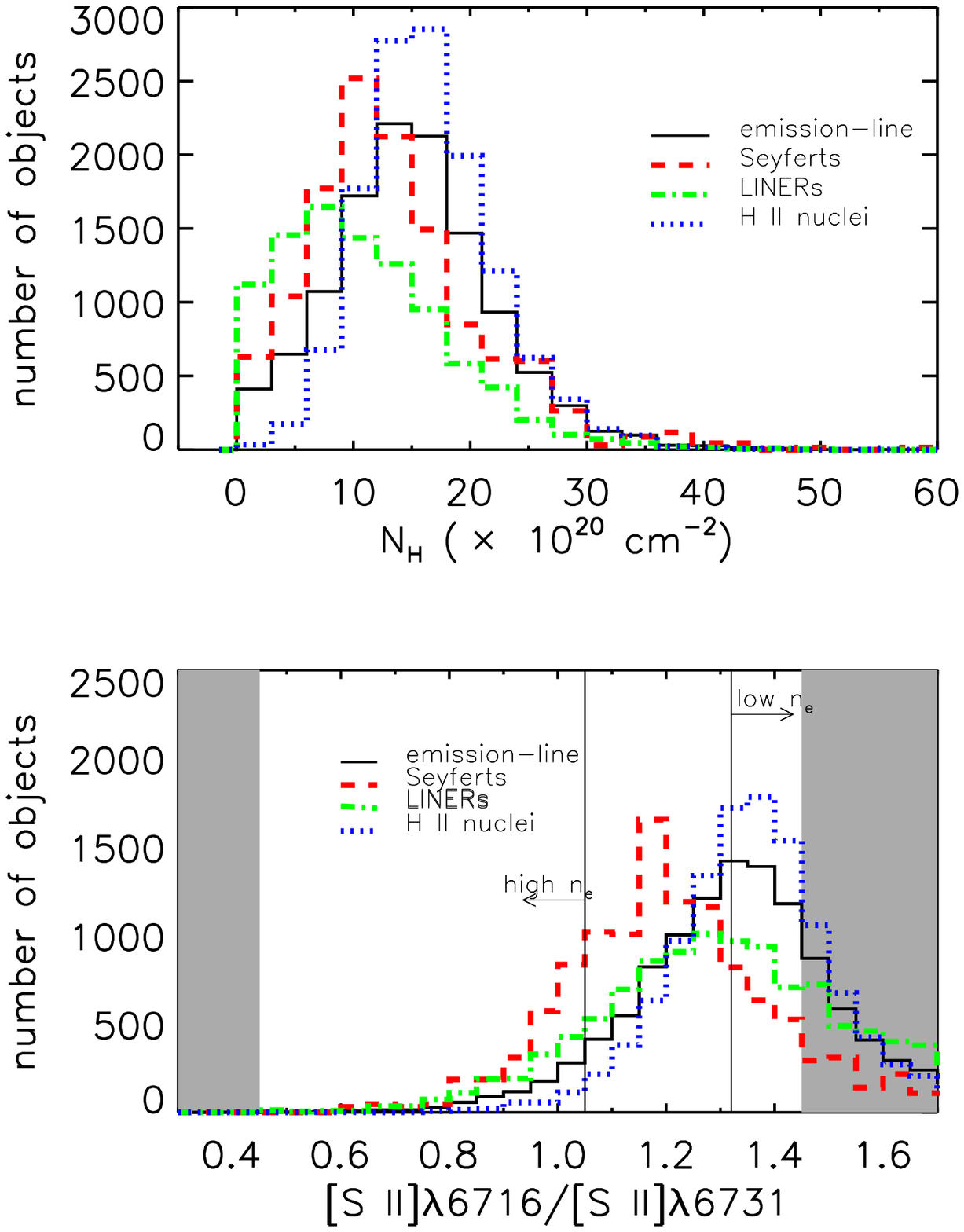}
\caption{ Distributions in log $L_{\rm [O I]}$, log $L_{\rm [O
  III]}$, intrinsic column density $N_H$, and
  [\ion{S}{2}]$\lambda$6716/$\lambda$6731 line ratio, for different
  subclasses of objects, in the absolute magnitude limited samples.
  In the latter plot, the shaded areas indicate corresponding values
  of $n_e$ that are beyond the minimum and maximum theoretical limits,
  max $n_e = 10^5$, min $n_e = 10^1$ cm.  Luminosities are in erg
  s$^{-1}$.  The histograms are re-normalized to have the total number
  of objects equal to that of the emission-line galaxy sample. 
\label{o1o3}}
\end{figure}

\begin{figure}
\plottwo{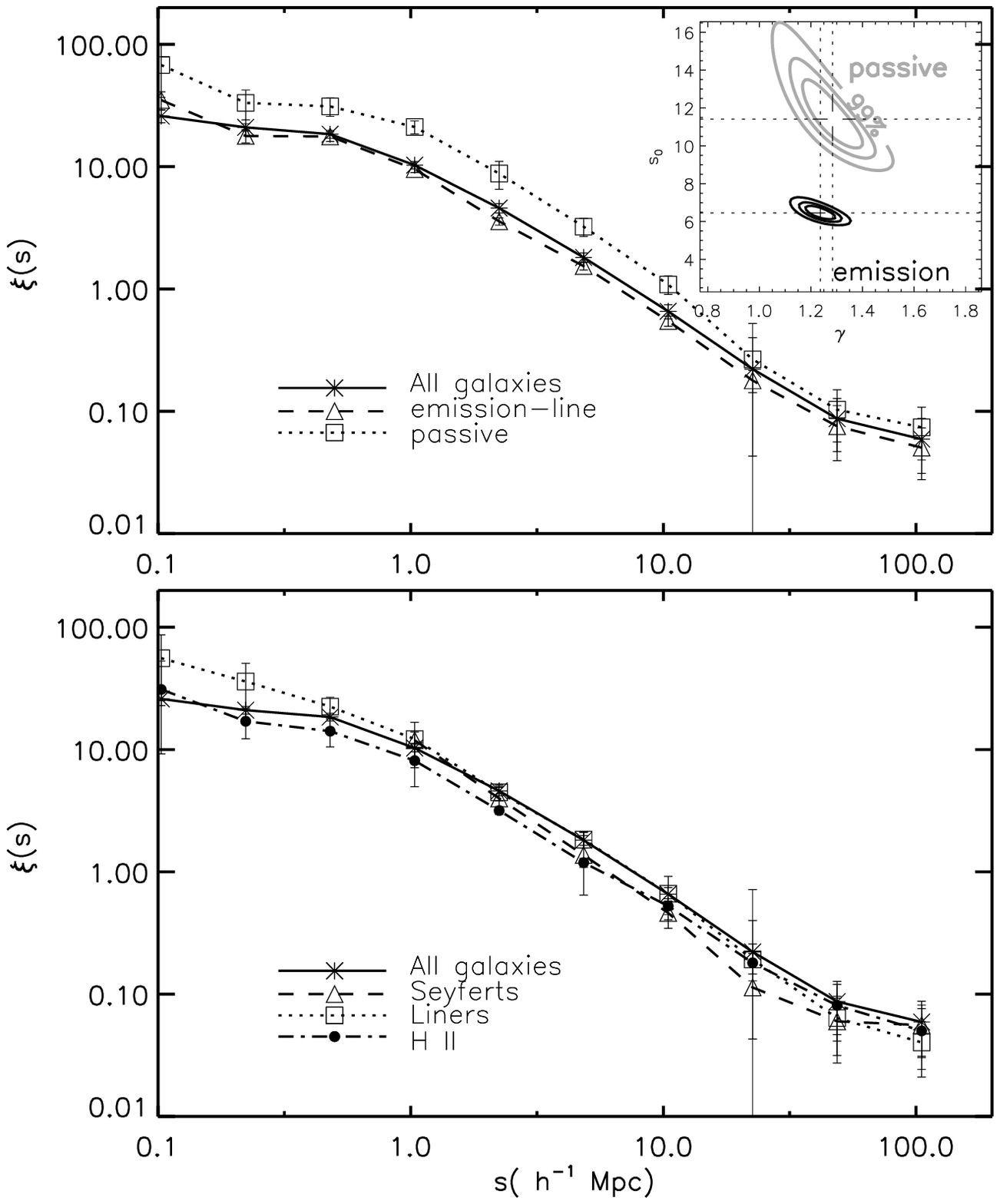}{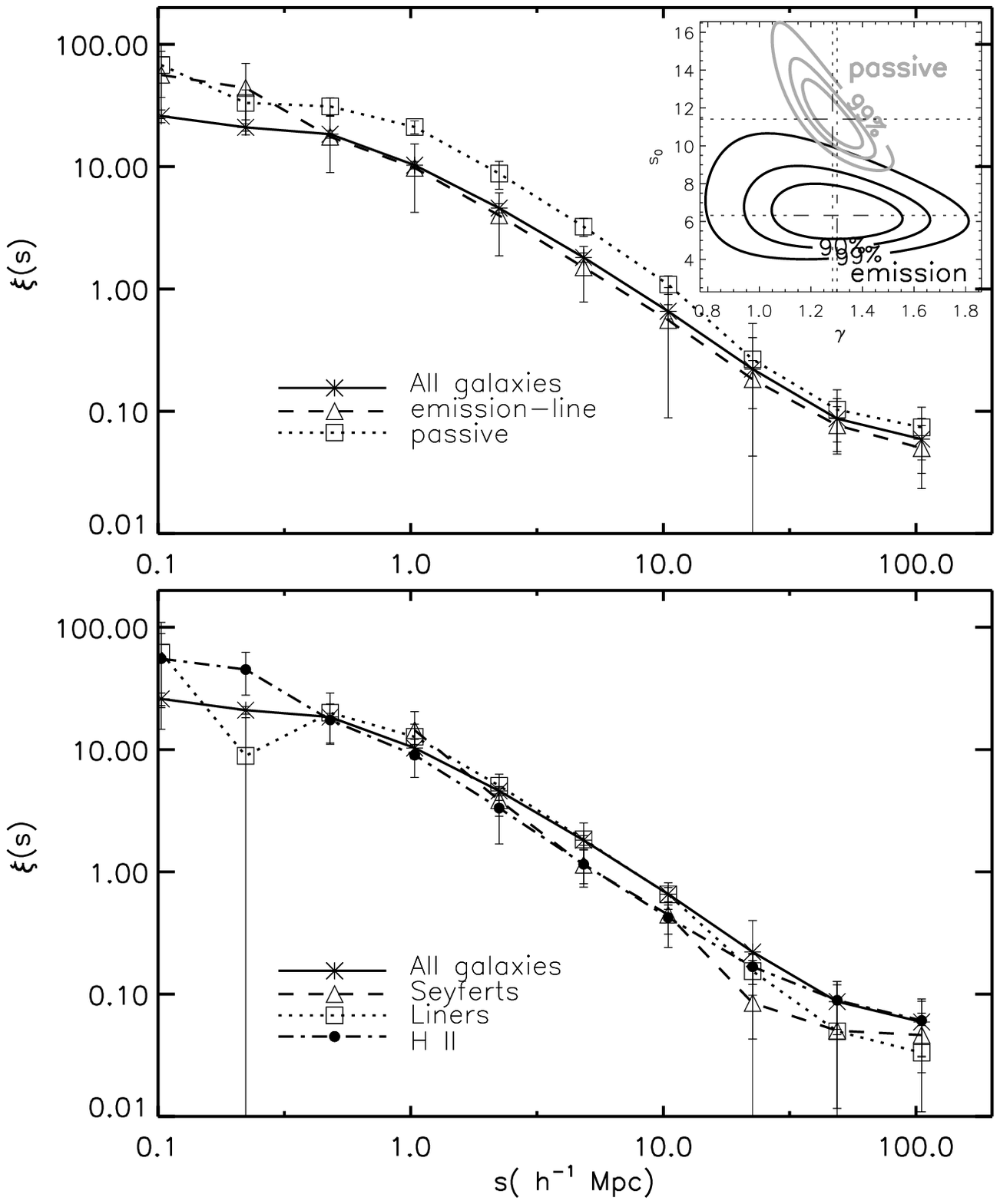}
\caption{ {\it Upper panels:} Comparison of the CFs estimated for
  the whole galaxy volume limited sample, for the objects that show
  activity in the form of emission lines, and for the ``passive''
  (i.e., relatively normal) galaxies.  {\it Lower panels:} Comparison
  of the CFs for different spectral classes of objects.  The insets
  in the upper plots show the $\chi^2$ contours indicating the
  68.3\%, 90\%, and 99\% confidence regions for the passive and the
  actively line-emitting galaxies.  The {\it left} panels correspond
  to spectral classification that excludes [\ion{O}{1}], while the
  {\it right} panels show the results for sample definition that
  includes this feature; here, the CF of the emission-line galaxies
  shows lower signal-to-noise and the power-law fits have larger error
  contours.  The passive galaxies are significantly more clustered
  than the ``active'' line-emitting galaxies.  $\chi^2$ contours
  corresponding to the CF's of Seyferts, LINERs and H {\sc ii}'s are
  illustrated in Figure ~\ref{csi_class}.
\label{tpcf_class}}
\end{figure}

\begin{figure}
\plottwo{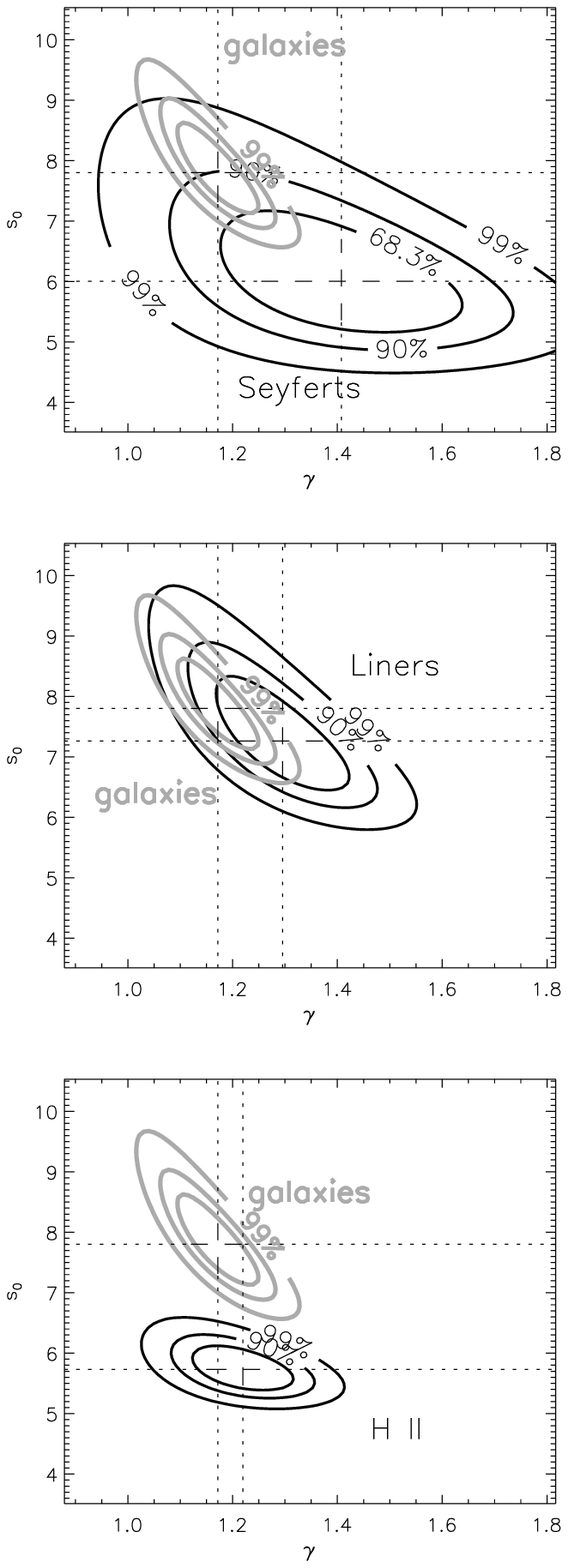}{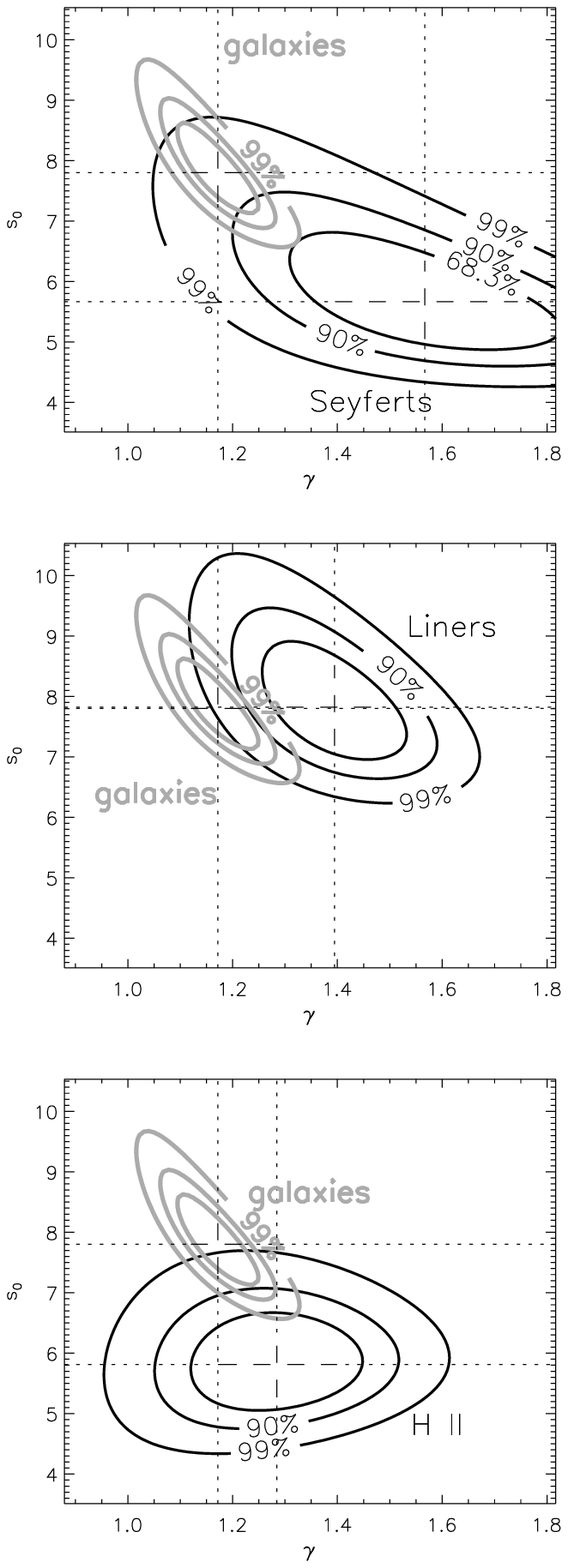}
\caption{ Confidence intervals of the power-law fit parameters, $s_0$ and
  $\gamma$. Contours indicate the two-dimensional 68.3\%, 90\%, and 99\%
  confidence regions for the CF of Seyferts, LINERs, and H {\sc ii}
  galaxies.  The {\it left panels} show results for sample definitions
  that do not employ conditions on [\ion{O}{1}], while for the {\it
  right panels} conditions on [\ion{O}{1}] are used.  Each plot shows,
  for comparison, the results for the volume limited galaxy sample;
  the axis range is the same in all plots.
\label{csi_class}}
\end{figure}

\begin{figure}
\plottwo{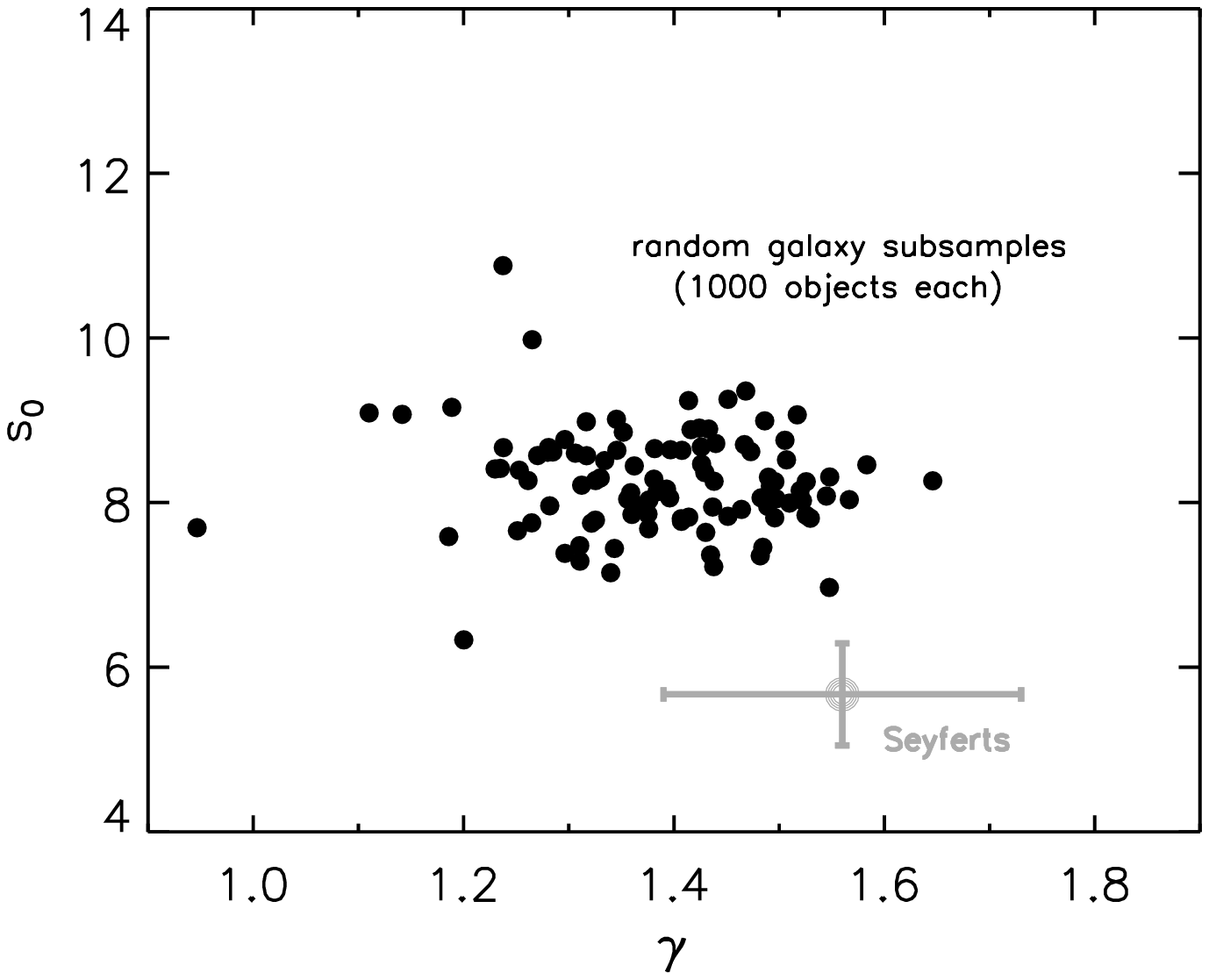}{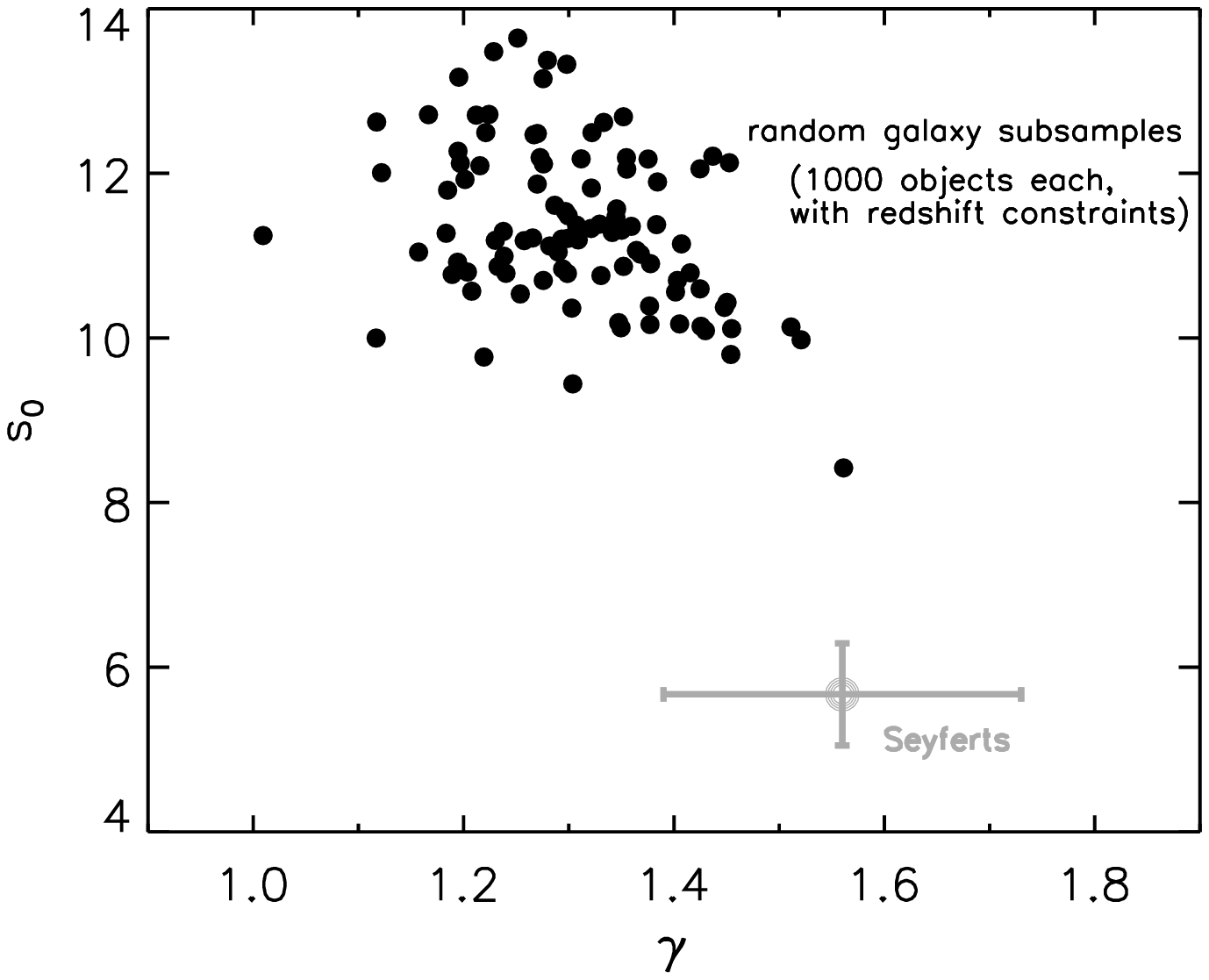}
\caption{Comparison of CF parameters $s_0$ and $\gamma$ for 100 random
  subsamples of the volume-limited galaxy sample with the CF for
  Seyferts whose definition includes [\ion{O}{1}].  The {\it left
  panel} shows the results for subsamples of galaxies constructed by
  randomly selecting 1000 objects per subsample, while the {\it right
  panel} shows the results for the case where the sources in the
  random subsamples are also constrained to follow the redshift
  distribution of the Seyferts.  Note that in both cases, not a single
  "mock Seyfert" sample has as low a value of $s_0$ as Seyferts
  display, and that in both parameters the real Seyferts are clearly
  separated from the mock samples.
\label{mock}}
\end{figure}

\begin{figure}
\plottwo{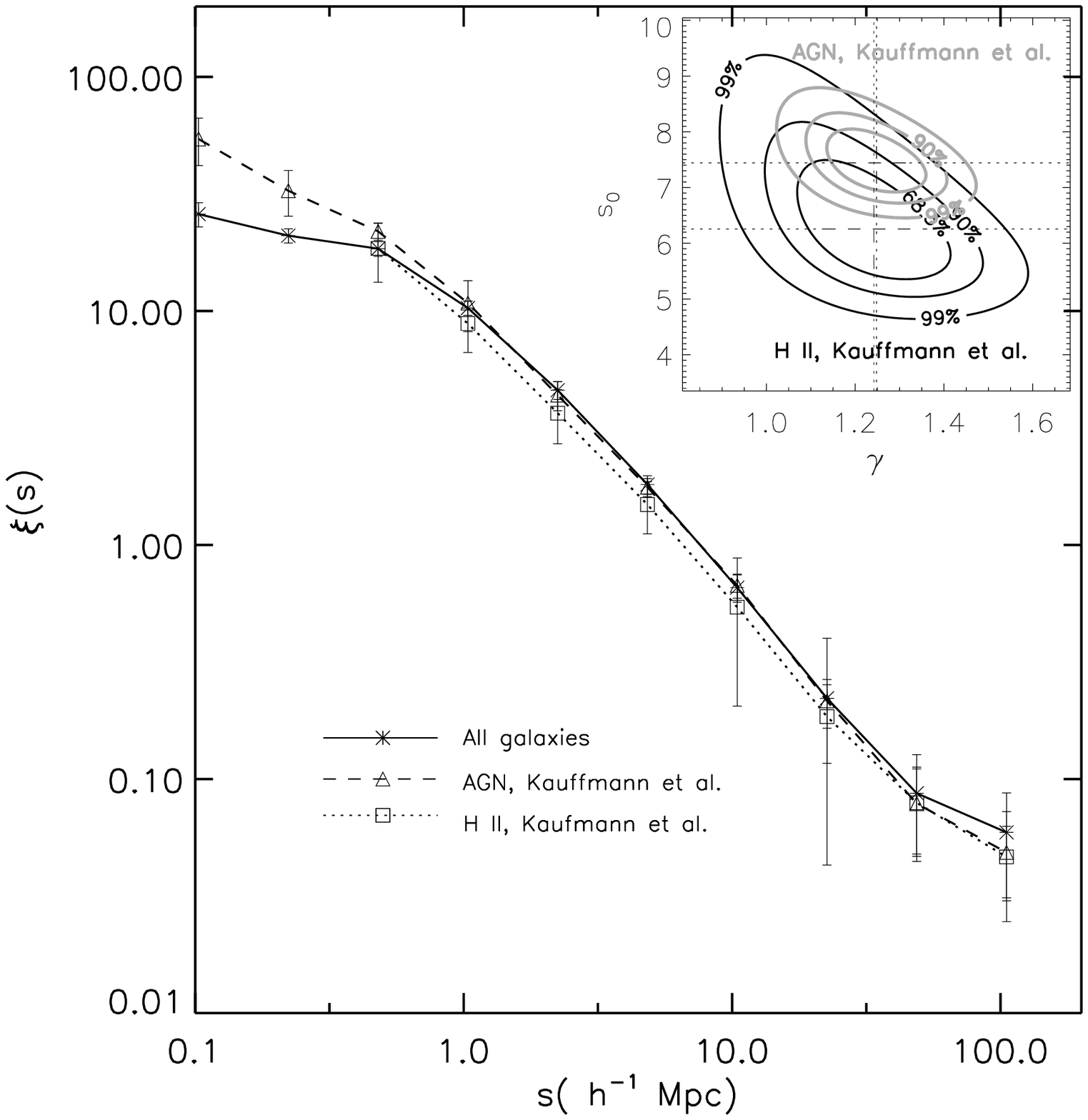}{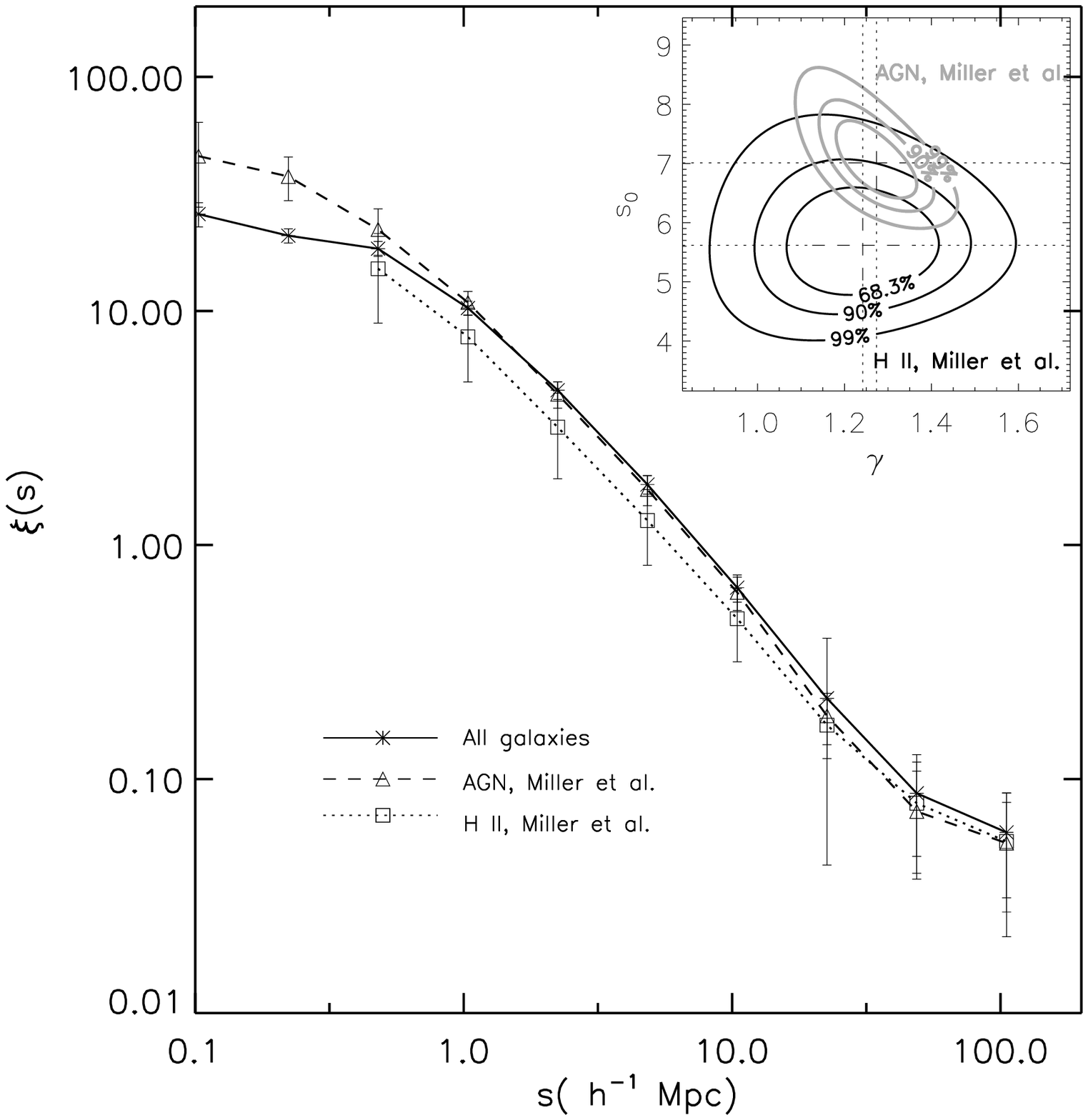}
\caption{ Comparison of the CFs estimated for AGN and line emitting
  non-AGN classified using the definitions of Kauffmann et al. ({\it
  left}), and Miller et al. ({\it right}).  When such AGN samples are
  analysed, the LINER behavior is dominant, thus, their clustering
  properties are similar to those of the galaxies.
\label{agnKM}}
\end{figure}

\begin{figure}
\plottwo{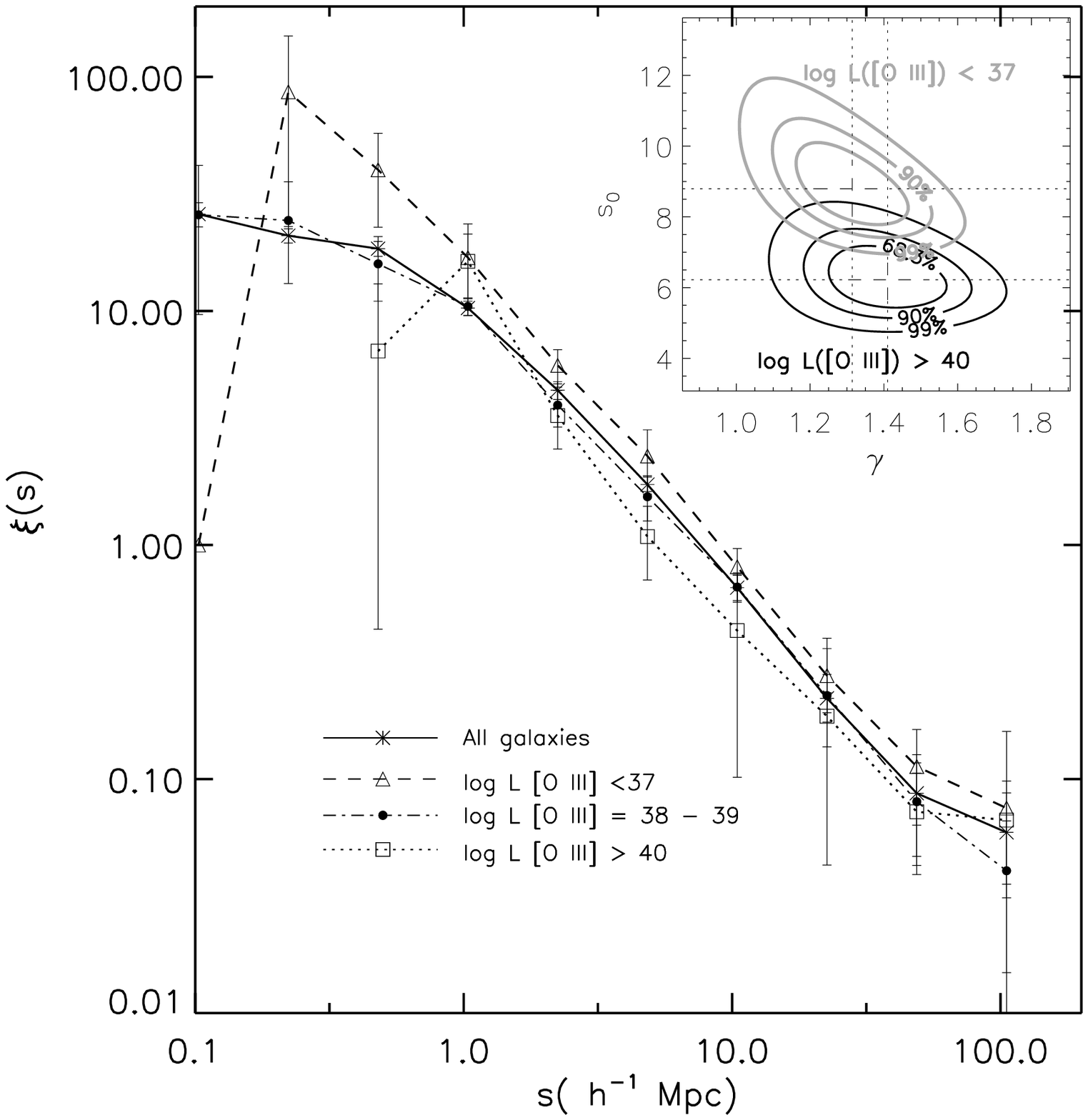}{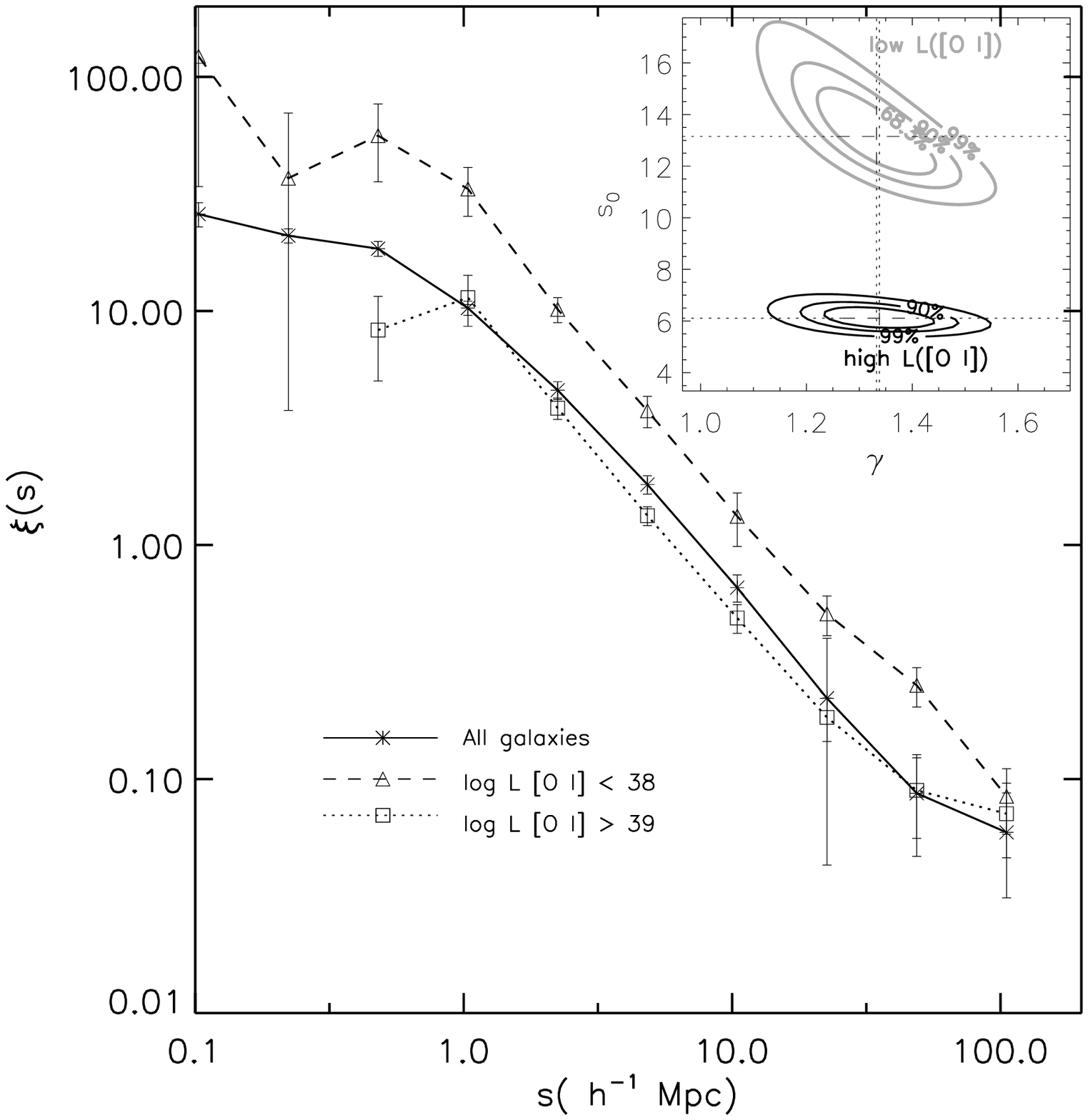}
\caption{ The CFs for object samples characterized by different
  ranges of [\ion{O}{3}] ({\it left}) and [\ion{O}{1}] ({\it right})
  emission-line luminosities.  The galaxy correlation function is also
  shown for comparison in both panels. Sources with lower $L_{\rm
  [O III]}$ and low $L_{\rm [O I]}$ show significantly stronger
  clustering.  Comparisons of the $\chi^2$ contours corresponding to the
  68.3\%, 90\%, and 99\% confidence regions are indicated in the inner
  upper right insets in each plot.
\label{tpcf_o1o3}}
\end{figure}

For objects with relatively high [\ion{N}{2}]/H$\alpha$ ratios (above
the Kewely et al. curve), [\ion{O}{3}] is, as discussed by
\citet{kau03}, only weakly affected by residual star-formation, and
therefore capable of distinguishing line emission activity due to
accretion from that originating in star formation.  As illustrated in
Figure ~\ref{o1o3}, using our spectral classification, the high
$L_{\rm [O III]}$ objects include most of the Seyferts.
Interestingly, the subset of luminous emission-line galaxies includes
almost no LINERs, which clearly dominate the low luminosity end of the
distribution.  This trend supports and justifies further the physics
behind the correlation found by \citet{kau03} between the ionization
state of the AGN, as expressed by their position angle in the
[\ion{N}{2}] diagnostic diagram, and the AGN [\ion{O}{3}] luminosity.
It is important to note that, because the number of Seyferts remains
small in general, H {\sc ii}'s remain the dominant type in the high
$L_{\rm [O III]}$ subset, comprising $\sim 40$\% of these objects; the
small fraction of H {\sc ii}'s with high $L_{\rm [O III]}$ makes a
large contribution to this subsample and so complicates interpretation
of our estimates of the CFs.  For the sake of comparison, note also
that strictly among AGN classified based on \citet{wak04} definition,
that implicitely exclude H {\sc ii}'s, the highest third in $L_{\rm [O
III]}$ remains still weakly represented by bona-fide AGN: the Seyferts
make up only 30\% of the objects, while the rest of them are
either LINERs or unclassified emission-line objects.

The CF of the high $L_{\rm [O III]}$ systems shows a low clustering
amplitude, as expected from the properties of the objects that
comprise the sample.  Both Seyferts and H {\sc ii}'s are weakly
clustered, with amplitudes that are similar, within the uncertainties.
The steep slope of this CF, $\gamma \approx 1.4$, suggests a strong
influence in behavior from the Seyferts.  The low $L_{\rm [O III]}$
sources exhibit a CF that is consistent in both the amplitude $s_0
\approx 8.8$ and the slope $\gamma \approx 1.3$ with what we measure
for the LINER sample.  Such strong correspondence may indicate that
low levels of activity, either originating in star formation or
accretion, is predominantly present in highly clustered hosts, and
thus more probable in crowded environments.

The variation in the clustering amplitude between the samples of low
and high $L_{\rm [O I]}$ is very strong, showing a difference of $\sim
6 \sigma$.  As with selection by $L_{\rm [O III]}$, the low $L_{\rm [O
I]}$ objects have large clustering amplitudes, while the more active
sources remain weakly clustered.  The dominant majority of low $L_{\rm
[O I]}$ objects are LINERs, suggesting again that galaxies that
manifest the lowest level of activity are the ones that are strongly
clustered.  LINERs remain very rare (less than 5\%) in the high
$L_{\rm [O I]}$ sample.  The high $L_{\rm [O I]}$ subset is highly
populated by H{\sc ii}'s ($\sim 60$\%), and, although this dataset
includes the majority of Seyferts, such systems make up only $\sim
9$\% of the objects.  Consequently, the clustering properties of the
high $L_{\rm [O I]}$ objects are close to those of Seyferts and
star-forming galaxies, revealing a relatively small clustering
amplitude.

As discussed in Sections \ref{samples} and \ref{prop}, a strong link
is expected between the strength of the [\ion{O}{1}] emission and that
of the accretion process.  Thus, the clustering dependence on $L_{\rm
[O I]}$ may be another indication of the relationship between the
level of AGN activity in galaxies and the properties of their host
dark matter halos: the active nuclei that exhibit the lowest levels of
activity due to accretion are the most clustered, and thus, the most
massive ones.  Any signature of star-formation activity is feeble, if
present at all in these systems; although numerically dominant among
the emission-line systems, the H {\sc ii}-like galaxies are
significantly underrepresented among the low luminosity objects (see
Figure ~\ref{o1o3}).  On the other hand, the most active sources,
whether in terms of star-formation or accretion, are weakly clustered,
thus we infer that these sources are hosted by relatively less massive
systems.

\subsubsection{Gas Density \& Intrinsic Extinction} \label{ne}

As Figure ~\ref{o1o3} illustrates, LINERs stand out among 
line-emitting objects in other aspects as well. 
Their electron densities are
lower than those of Seyferts but generally higher than in H {\sc ii}s,
and they show little intrinsic obscuration.  A clear segregation of
LINERs and Seyferts as a function of these parameters, similar to that
as a function of $L_{\rm [O I]}$ and $L_{\rm [O III]}$, is 
possible only for very small subsamples, and thus CF estimates
remain impractical.  Nevertheless, an investigation of the degree to
which the clustering amplitude depends on $n_e$ and $N_H$ remains of
interest even if the subsamples characterized by low and high values
of these parameters comprise a mix of spectral types.  Such analysis
offers additional insights into the origin of the clustering
differences exhibited by sources with contrasting physical conditions,
especially those that may characterize the availability of fuel for
accretion or star-formation.

\begin{figure}
\plottwo{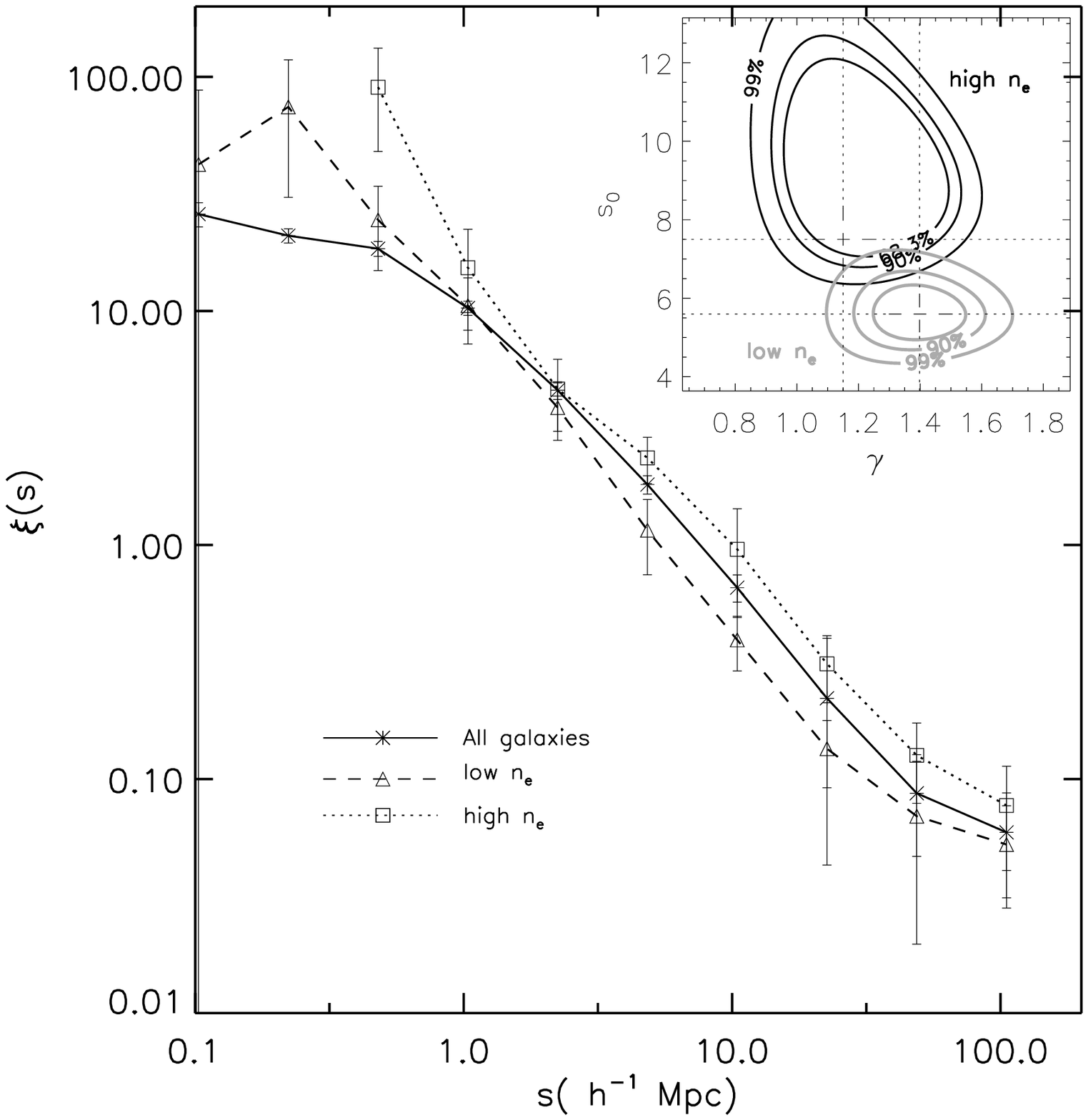}{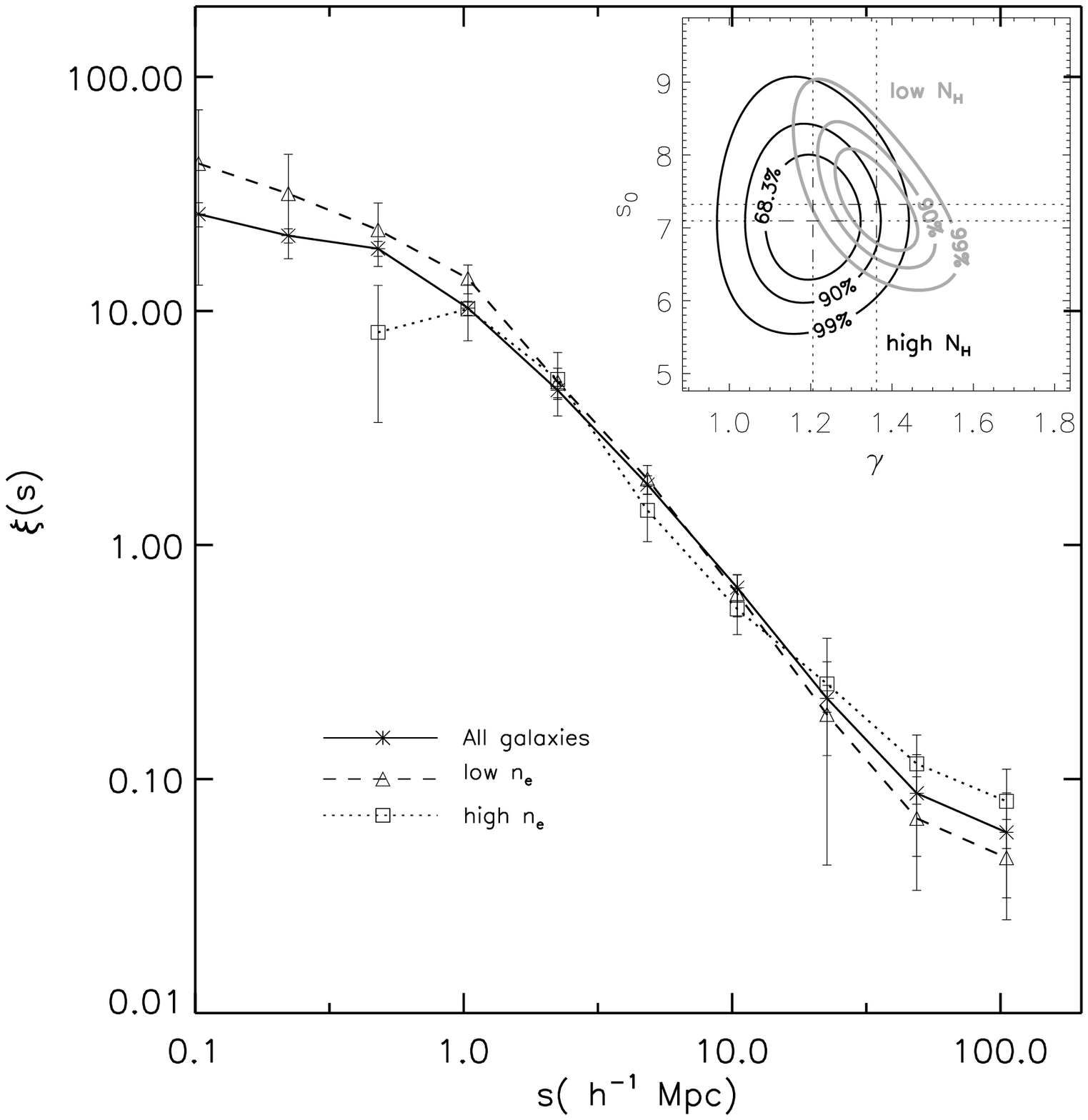}
\caption{ {\it left:} CFs for object samples that are characterized by
  different electron densities ($n_e$) as obtained from [\ion{S}{2}]
  line ratios (see text for details).  {\it right:} CFs for object
  samples with high and low levels of intrinsic absorption, $N_H$.
  The galaxy correlation function is also shown for comparison in each
  plot.  Confidence limits for the power-law fits to the CF for the
  subsamples of low and high $n_e$ and $N_H$ respectively are compared
  and shown in the upper right inset of each panel.
\label{tpcf_nenh}}
\end{figure}

Figure ~\ref{tpcf_nenh} shows a comparison of the CFs estimated for
subsamples characterized by low and high values of $n_e$ and $N_H$,
with the sample definitions given in Tables ~\ref{tbl-ne} and
~\ref{tbl-nh}.  Consistent with the trends found so far, the low $n_e$
subsample, in which 70\% of the objects are H {\sc ii} galaxies, has a
significantly lower clustering amplitude.  Interestingly, the high
$n_e$ subsample shows a high clustering amplitude that is consistent
with that of the LINERs and the average galaxy sample even though it
comprises a generously mixed population of emission line systems: 40\%
LINERs, 18\% Seyferts, 19\% H{\sc ii}s, and 23\% unclassified
line-emitters (that do not comply simultaneously with all the three
diagnostic diagrams, see Section ~\ref{samples}).  It is striking that
only a slight number dominance of the LINERs in this subsample of
sources with relatively high $n_e$ leads to such results.  This trend
suggests that the moderately high $n_e$ unclassified emission-line
objects have clustering properties more similar to LINERs than to
Seyferts or H{\sc ii}s.

The subsamples split on $N_H$ shows clustering trends that are
consistent with the results presented so far.  There is a slight
suggestion that the less obscured sources ($N_H < 10 \times 10^{20}$
cm$^{-2}$), in which the LINERs are in majority, are more clustered,
while the highly obscured systems, which are mostly star-forming
objects, are less clustered.  These differences reflect once more a
potential link between the amount, nature, and availability of the
fuel that powers the galactic activity and their clustering
characteristics: the dusty spirals that usually live in less crowded
environments are more likely to harbor H {\sc ii}'s in their centers,
while the nuclei of ellipticals would be generally dust-free due to
ram-pressure stripping in their more crowded, cluster-like habitats,
and thus would show LINER-like activity and physical properties.

\subsection{CF Estimates and Host Galaxy Properties} \label{host}

Studies based on large redshift surveys, 2dF and SDSS in particular,
have clearly shown that spatial clustering of galaxies depends on
galaxy structure (or morphology), luminosity, and internal star
formation history, as measured by spectral type or colors
~\citep{nor01, nor02, zeh02, zeh05, bud03, mad03}. The sense of this
dependence is that more luminous, redder, and earlier type galaxies,
which also harbor older stellar populations, are generally more
clustered than the rest of the galaxies.

\begin{figure}
\plottwo{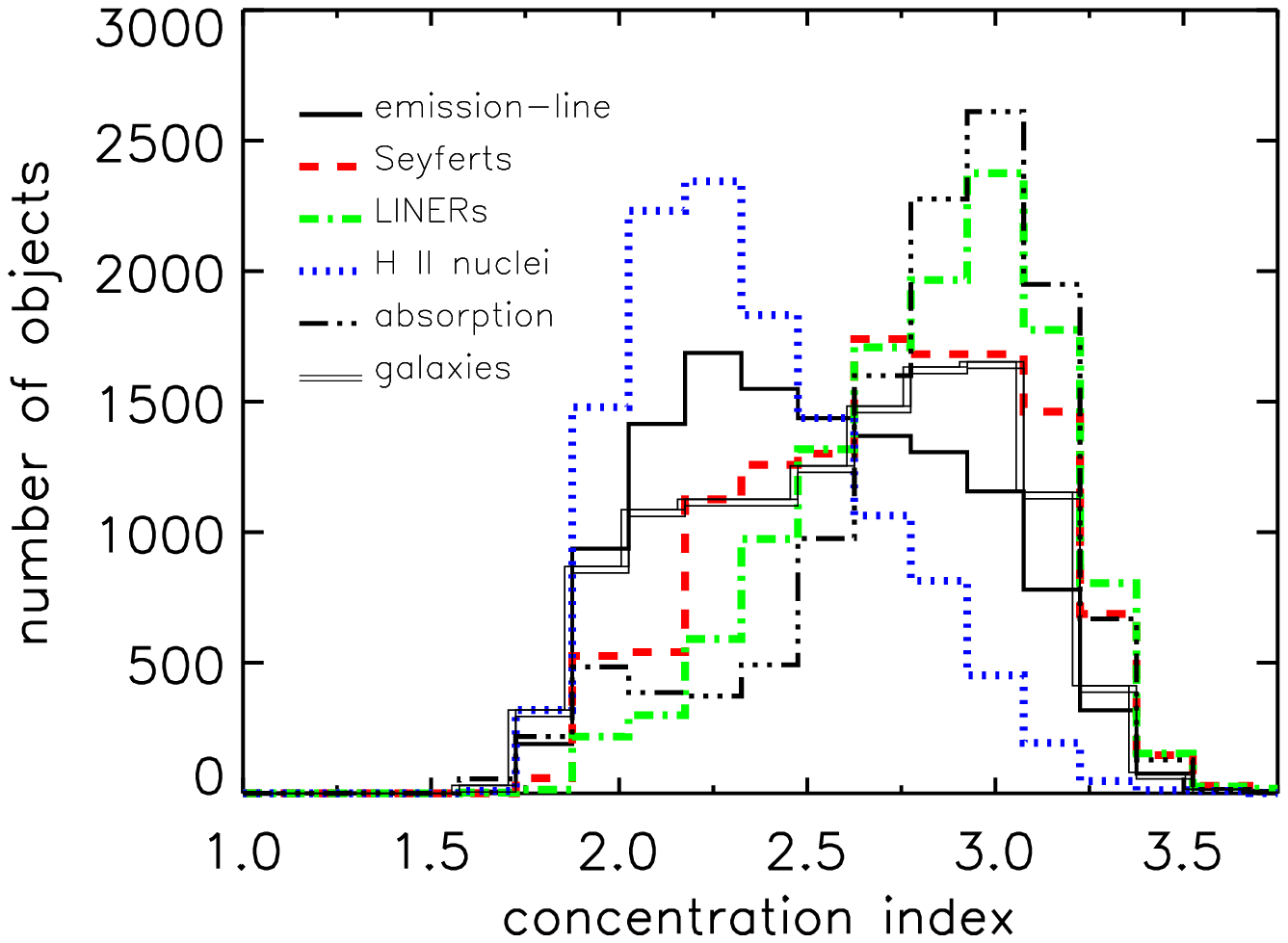}{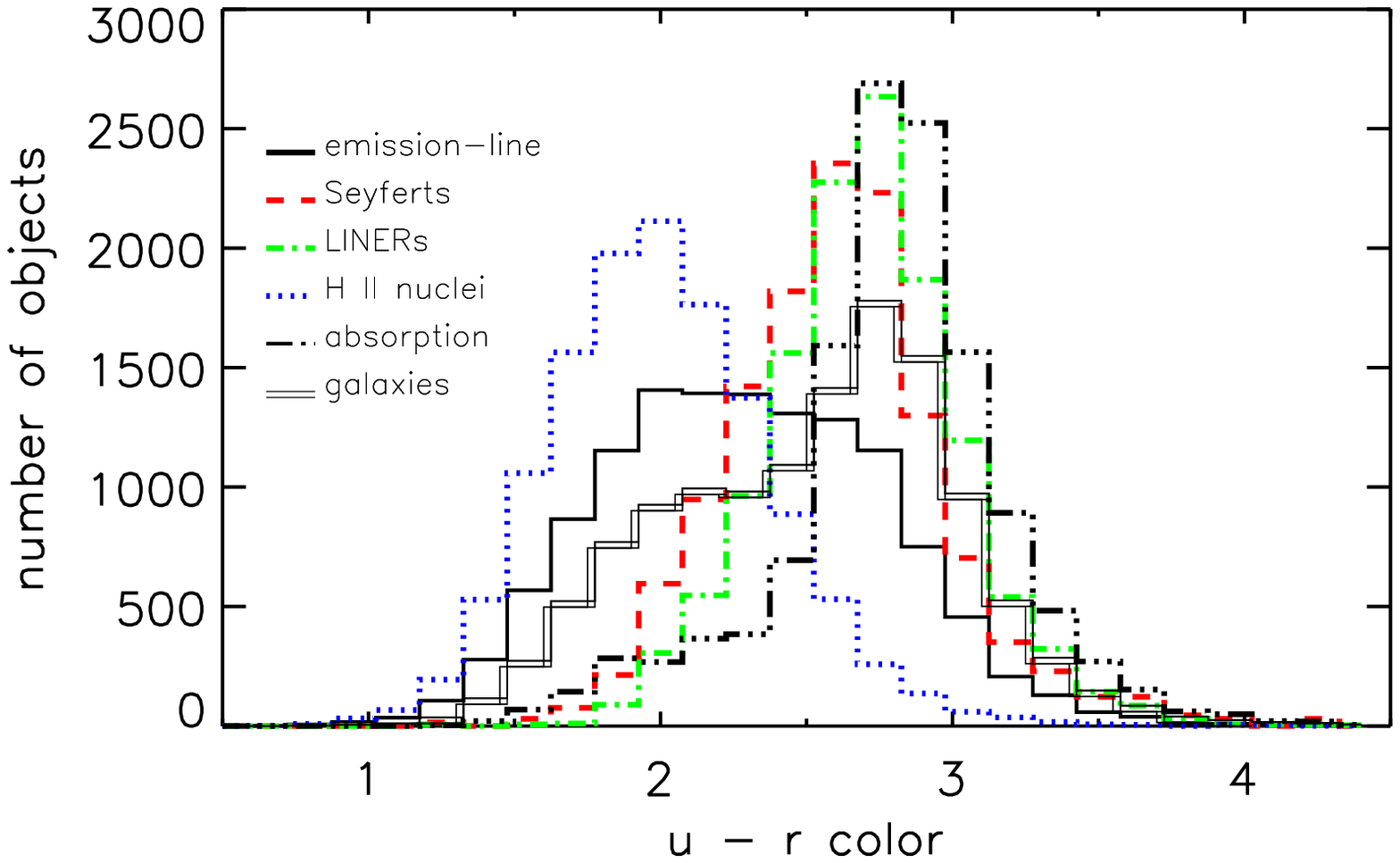}
\caption{ Distribution in the concentration index $C$ ({\it
  left}), and the $u-r$ color ({\it right}) for different subclasses
  of objects, in the absolute magnitude limited samples.  The
  histograms are re-normalized as in Figure ~\ref{o1o3}.  Note the
  similarity between the distributions in these parameters for the
  Seyfert and LINER samples.
\label{conc}}
\end{figure}

The well-described dependence of clustering on galaxy properties might
explain some of the difference we see between the clustering amplitude
of the H {\sc ii}'s, which seem to prefer bluer, later-type hosts, and
that of AGN-like objects (i.e., Seyferts and LINERs), which are more
likely to be hosted by red, early type galaxies.  As illustrated in
Section ~\ref{samples}, AGN samples defined based on simple
classification schemes are dominated by LINER-like behavior and are
therefore more clustered than the H {\sc ii} systems (i.e., Figure
~\ref{agnKM}). The median values of concentration indices, colors, and
other parameters of the host galaxies of our
spectroscopically-classified subsamples (see Figure ~\ref{conc}, and
Table ~\ref{tbl-param}) clearly illustrate the distinct preferences of
the stellar and non-stellar photo-ionized emission-line activity for
different host types.

The variation of clustering among Seyferts, LINERs, and H {\sc ii}
systems is not, however, so easily explained. When we use more
detailed spectroscopic classification to separate Seyferts and LINERs,
we find that they differ significantly in their clustering properties
despite our finding, consistent with previous studies \citep{ho97b},
that the large-scale, global properties of their hosts are quite
similar.  Although there is a slight shift of the Seyferts toward
bluer and later type hosts than LINERs, comparison using a
Kolmogorov-Smirnov (KS) test shows that their distributions in
concentration index, $C$, and color $u-r$ are consistent with being
withdrawn from the same parent population (KS$_{S-L}^{~~C}$ = 0.104,
KS$_{S-L}^{~~u-r}$ = 0.617).  LINERs' host morphology ($C$) and color
($u-r$) distributions are quite different from those of the whole
galaxy sample, both in the shapes and median values of these
distributions (KS$_{gal-L}^{~~C}$ = 4.8e-14, KS$_{gal-L}^{~~u-r}$ =
5.6e-11).  Yet, the clustering amplitudes of LINERs and typical
galaxies are very similar.  Likewise, Seyferts and the passive
galaxies are close to sharing the same parent distributions in host
properties (KS$_{Passive-S}^{~~C}$ = 0.108, KS$_{Passive-S}^{~~u-r}$ =
4.2e-4), although their clustering properties are markedly different.
Furthermore, the clustering amplitudes of Seyferts and H {\sc ii}'s
are quite similar, but their galactic hosts are, beyond any doubt,
very different in type.

To summarize, using broad photometric morphological measures (e.g.,
concentration index and color), we find no evidence that the
difference in clustering amplitudes of different spectral classes of
galaxies, and in particular Seyferts and LINERs, is driven by the
density-morphology relation.

\begin{figure}
\plottwo{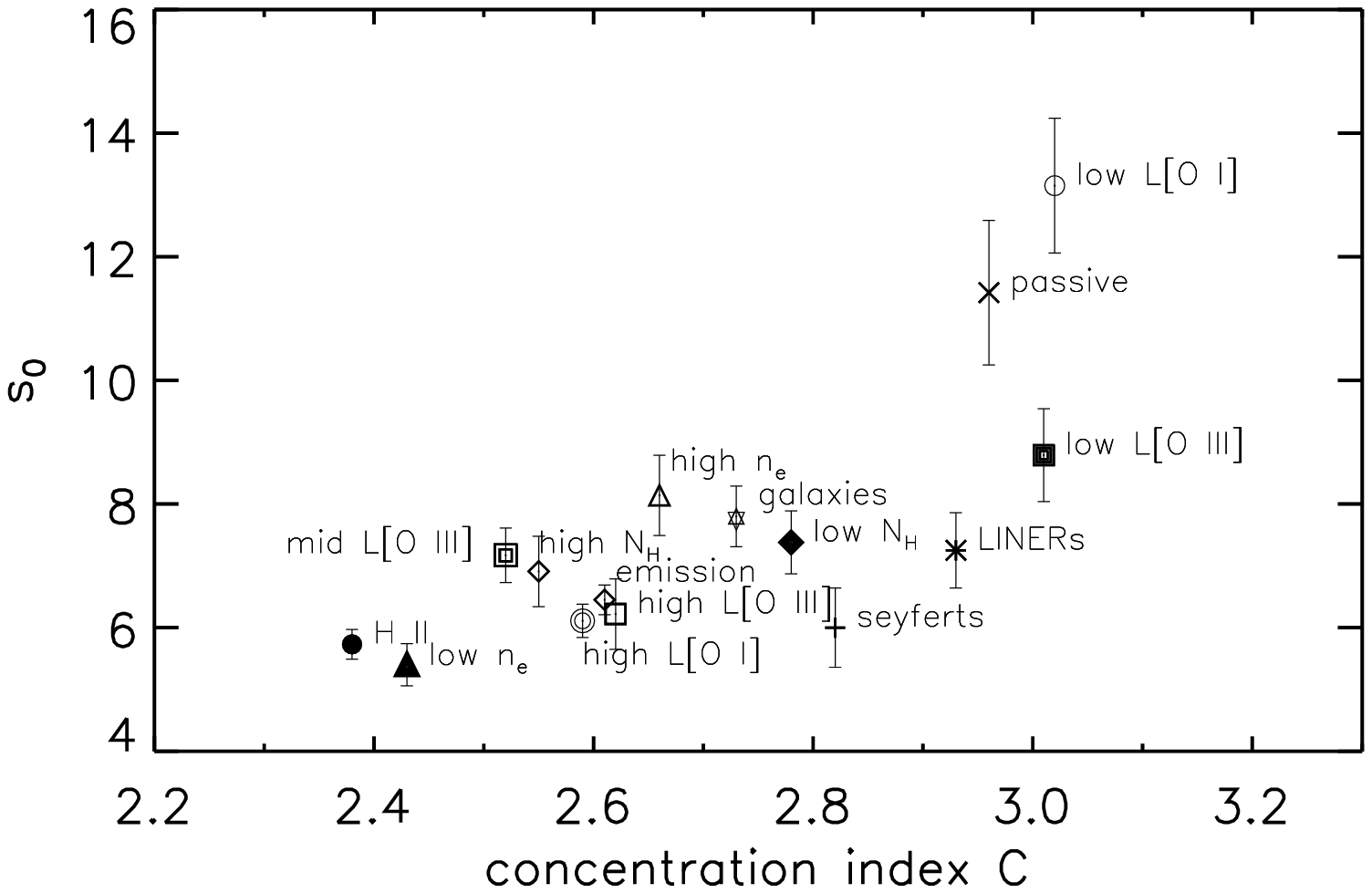}{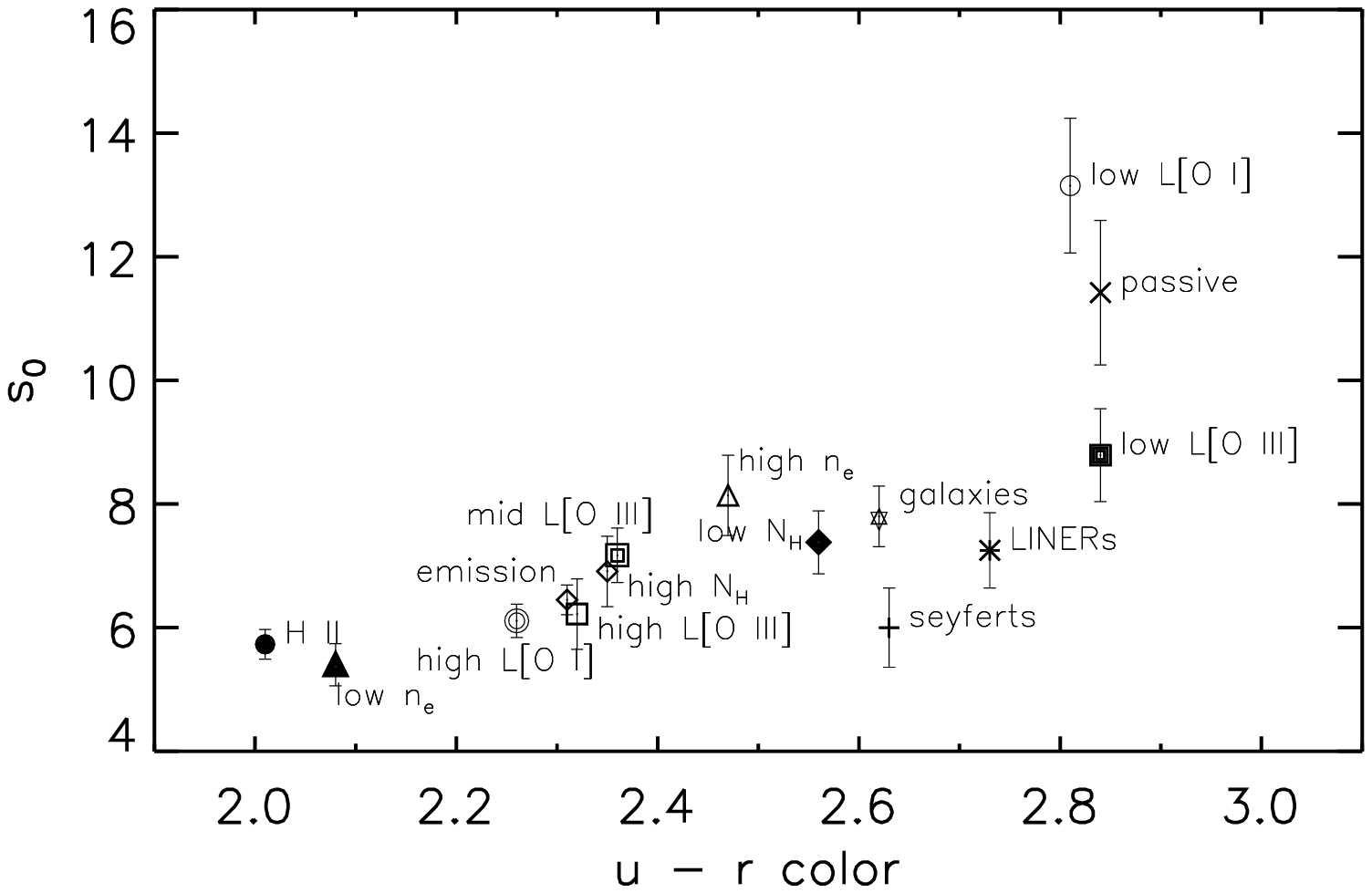}
\caption{ Relationships between the clustering amplitude $s_0$ and
  the median values of the concentration index $C$ ({\it left}), and
  the $u-r$ color ({\it right}), for the corresponding samples of
  different emission-line properties.  
\label{rho-c}}
\end{figure}

In the light of the effects the host characteristics have on
clustering, the differences in $s_0$ exhibited by the low and high
$L_{\rm [O I]}$ and $L_{\rm [O III]}$ samples merit further attention
as well.  As it can be seen from Figure ~\ref{rho-c}, these datasets
are clearly distinguishable in their median concentration indices and
$u-r$ colors, with which the clustering amplitudes seem to correlate.
The sources with low [\ion{O}{1}] and [\ion{O}{3}] luminosities are
also the reddest $u-r \approx 2.8$ and generally early in their
morphological type, as revealed by their high median value of $C
\approx 3$, and present strong clustering.  The most luminous sources
in [\ion{O}{1}] and [\ion{O}{3}] are bluer ($u-r \approx 2.3$) and of
later type morphologies ($C \approx 2.6$), and are weakly clustered.
Thus, it may be apparent that the different clustering amplitudes for
the high and low luminosity objects is at least partially driven by
the morphology-density relation that galaxies usually obey.  However,
based on exhaustive studies of the color and host type dependence of
the correlation function, ~\citet{zeh02, zeh05} reveal differences in
$s_0$ of only $ ^{<}_{\sim} 2 h^{-1}$ Mpc, which are smaller than
those between our low and high $L_{\rm [O III]}$ systems, and
especially those we observed between the objects with high and low
$L_{\rm [O I]}$ (see Table ~\ref{tbl-1}).  Moreover, our subsamples
span a significantly reduced range of values in their median $C$ and
$u-r$ than those employed in the Zehavi et al. comparisons, suggesting
that, the morphology-color effect on the discrepancy in $s_0$ measured
in our samples should be even smaller.  We thus conclude that the
differences in the clustering amplitudes of objects that differentiate
themselves through their [\ion{O}{1}] and [\ion{O}{3}] luminosities
are primarily constrained by these parameters, and only secondarily by
their host properties. In the following section we discuss various
scenarios that may explain such effects.


\section{Discussion} \label{discussion}

\subsection{AGN Fueling \& Life Cycle}


Under simple assumptions, the amplitude of the correlation function
can be used to estimate the typical mass of the dark matter halos in
which the objects reside \citep{kai84, gra04, mag04}, and for AGN in
particular, to estimate their typical lifetime \citep{mar01}.  If, as
expected in the hierarchical picture of structure formation, the most
massive systems are more clustered, our results suggest that Seyferts
live in less massive halos than those that host LINERs.  The empirical
relation between black hole mass $M_{\rm BH}$ and galaxy mass $M_{\rm
DM}$ \citep{fer02, bae03}, that is based on the $M_{\rm BH} -\sigma_*$
relation which seems to hold similarly for both inactive galaxies
\citep{geb00, fer00} and AGN hosts \citep{geb00b, fer01, nel04, onk04,
gre05}, implies that the black holes harbored by these galaxies
must scale in the same manner. Thus, Seyferts' black holes must be
smaller than those that live within LINERs.  Via direct comparisons of
2PCF for simulated dark matter and for observed SDSS galaxies,
\citet{wak04} reached similar conclusions regarding the masses of the
black holes.


Together with the different trends we see in the physical properties
of active galaxies with distinct spectral classification, this
possible distinction in their BH mass offers the missing link in our
understanding of their intrinsic nature.  The high $L_{\rm [O I]}$ and
$L_{\rm [O III]}$ of Seyferts suggest that their smaller BHs are
accreting at relatively higher rates or maybe more efficient than the
larger BHs whose activity is classified as LINER-like.  The low
activity level of LINERs' massive BHs may well be a consequence of the
fact that these systems have less fuel available for accretion.  Both
the electron density and column density of surrounding (dusty)
material are significantly lower in these objects, suggesting a
scarcity of matter supply.  Seyferts, on the other hand, show the
highest values of $n_e$ and a relatively large range in $N_H$
suggesting that in these sources fuel can easily be provided at levels
sufficient to sustain accretion at modest levels.

In this scenario, Seyferts grow their BHs at a relatively fast
pace, and this activity would be expected to last for short periods of
time.  Such a highly active phase would end with the exhaustion of
surrounding accretion fuel, when the BH mass reaches levels
similar to those of LINERs, after which weak (or inefficient) accretion can
continue for much longer time.  This picture is consistent with the relative 
numbers of Seyferts and LINERs.  If Seyferts are simply the
manifestation of the stronger but short accretion phase, then their
numbers must be small, especially when compared to those of LINERs
that subsist much longer at low activity levels and thus are more common 
among active galaxies.  The results of our AGN
classification, together with other previous studies discussed above
(see section ~\ref{samples}) indicate indeed that LINERs
numerically dominate the AGN samples, as the LINER-type emission is
found in at least three times more galaxies than Seyfert-like activity.

If LINERs harbor more massive black holes then, as suggested by the
close connection between BH mass and bulge size in their host galaxies
\citep{mag98, gra01}, LINERs' hosts must be more bulge-dominated than
those of Seyferts.  Although we have seen that the Seyfert and LINER
overall distributions in $C$ and color are very similar, the median
values show small differences (see Table ~\ref{tbl-param}) indicating
slightly higher $C$ and redder $u-r$ colors for LINERs, consistent
with the expected trend.  Future larger SDSS samples, together with
more successful morphology classification schemes \citep{par05b}
should allow more detailed investigations of such feature.

\subsection{The H {\sc ii} \& the Transition Systems}

Further, we may extrapolate this picture to galaxies whose nuclear
activity is dominated by star-formation, the H {\sc ii}
systems. Assuming that the same scaling relations hold for these
sources, the weak clustering of H {\sc ii}'s implies that they harbor
relatively small black holes in their centers.  In general, they show
moderately low, quiescent [\ion{O}{1}] and [\ion{O}{3}] luminosities.
However, because they are very numerous, the high luminosity tail of
their line-emission distribution dominates the statistics of the
sub-sample of luminous actively line-emitting galaxies.  There is
a generally high level of obscuration in these objects, as shown
by their large column densities $N_H$.  The overall low $n_e$ values
that characterize the emitting gas in H {\sc ii}'s suggest that fuel
for eventual accretion is tenuous in these galaxies.
Thus, if their BHs are accreting, the rates must not be high or, if
the BHs are actively accreting, the the resulting radiation is highly
obscured.  In other words, even if the kinetic energy output from the
central few tens of parsecs is as strong as that of the Seyferts,
there is only a very small fraction that gets transferred to the
thermal plasma to produce the enhancement in the optically
collisionally excited lines, at the level seen in the
accretion-powered Seyferts or LINERs.  Consequently, any accretion
signature due to robust activity would be sub-dominant in the H {\sc
ii}'s spectra because it would be buried in the surrounding emission
originating in stellar heating.

\begin{figure}
\plottwo{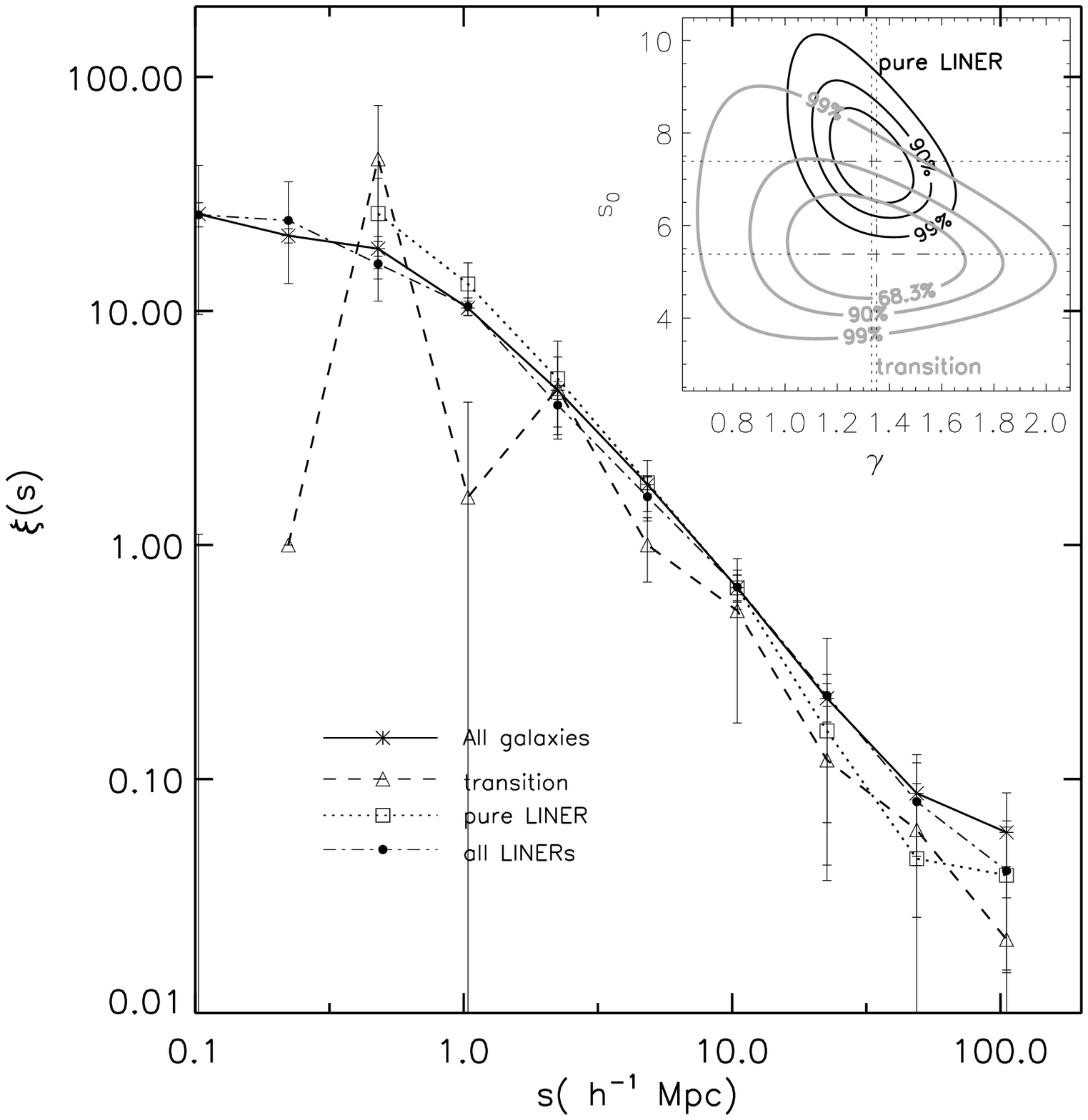}{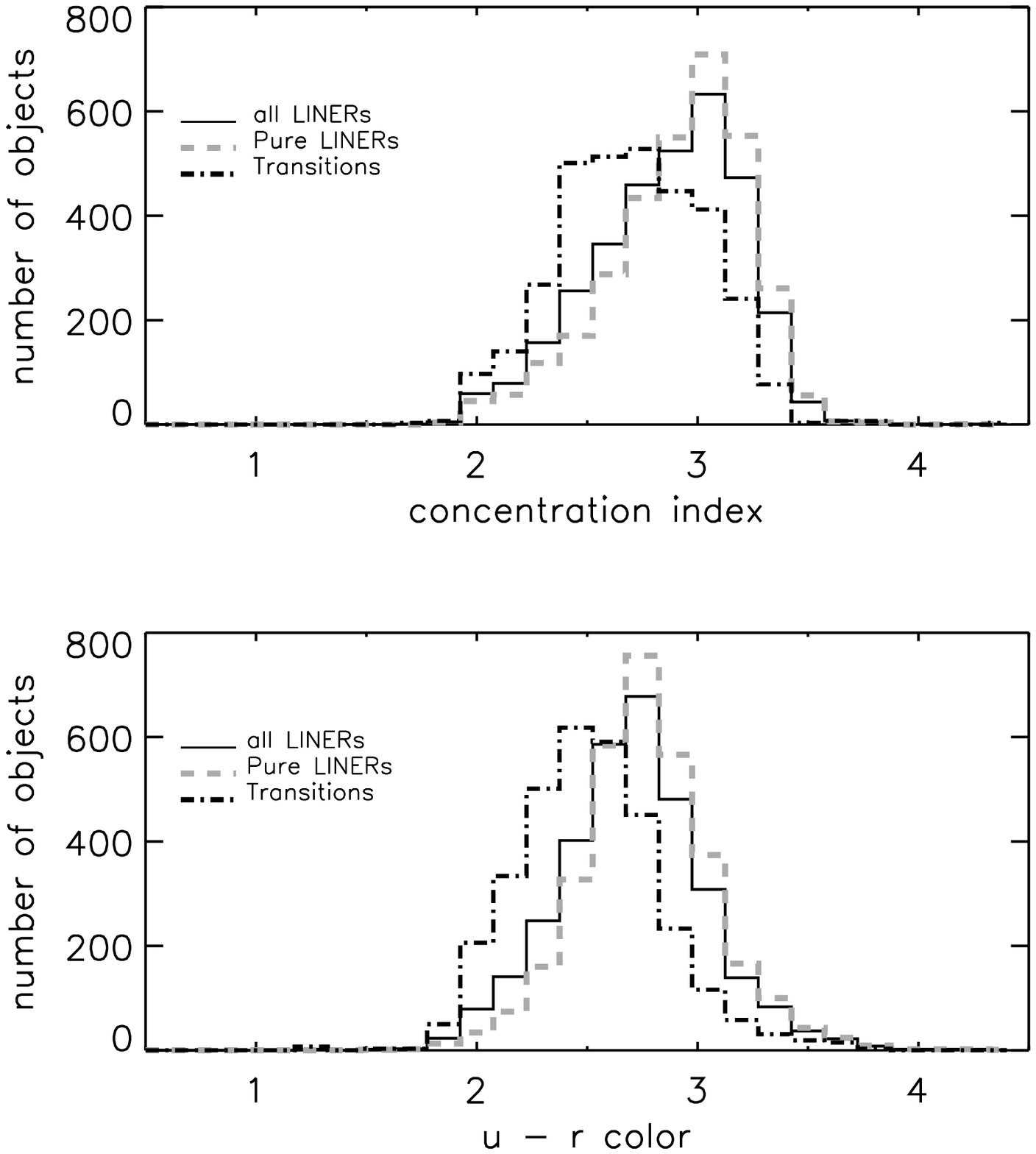}
\caption{ {\it left:} CFs corresponding to Pure LINERs and
Transition objects: the transition systems are significantly less
clustered than the Pure LINERs. {\it right:} Comparison of
distributions in $C$ and $u-r$ for these two subsets of LINERs; the
histograms are re-normalized to have the total number of objects equal
to that of the whole LINER sample.  Note the slight shift toward bluer
and later type hosts for the Transition objects.
\label{transition}}
\end{figure}

\begin{figure}
\plottwo{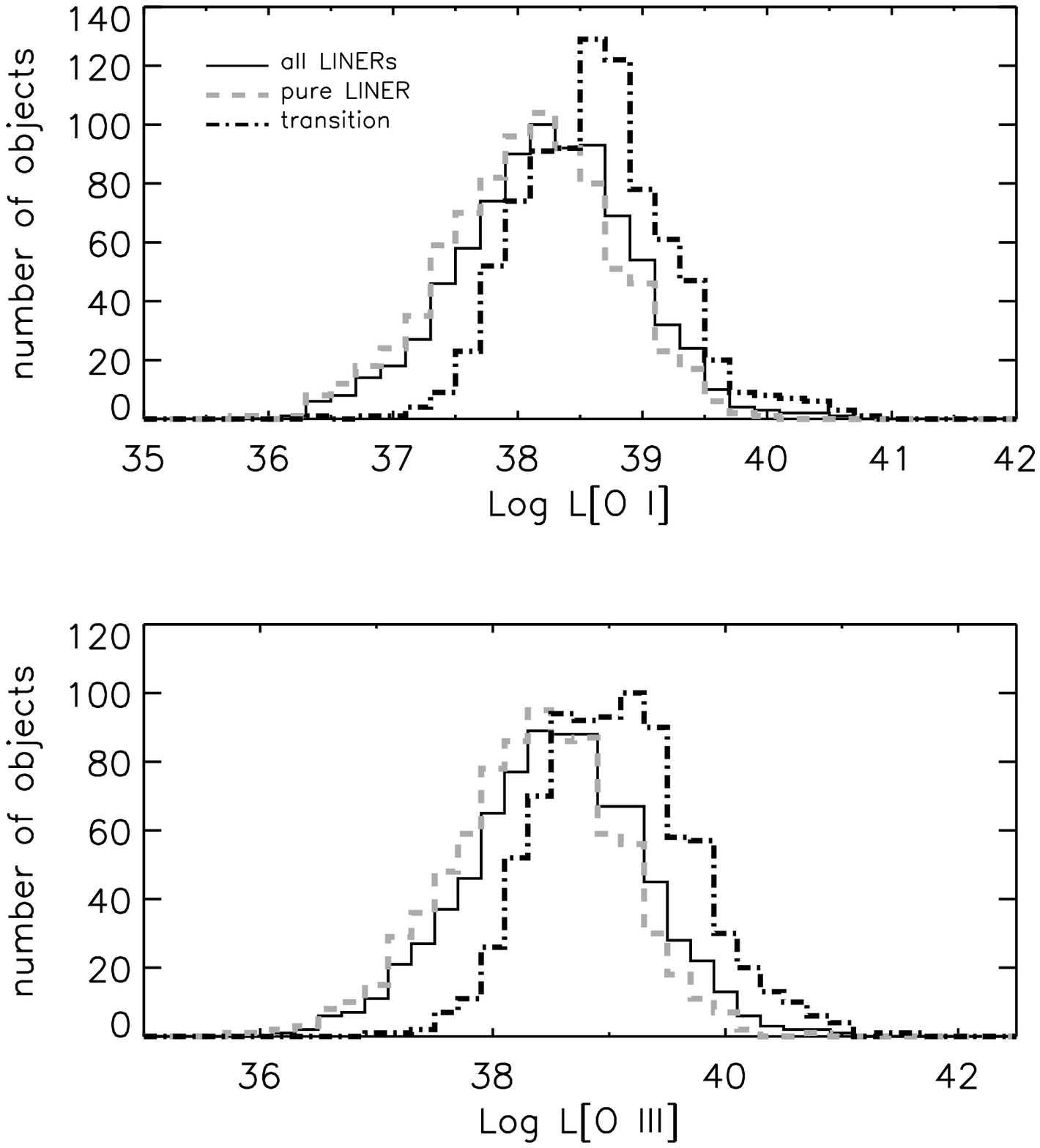}{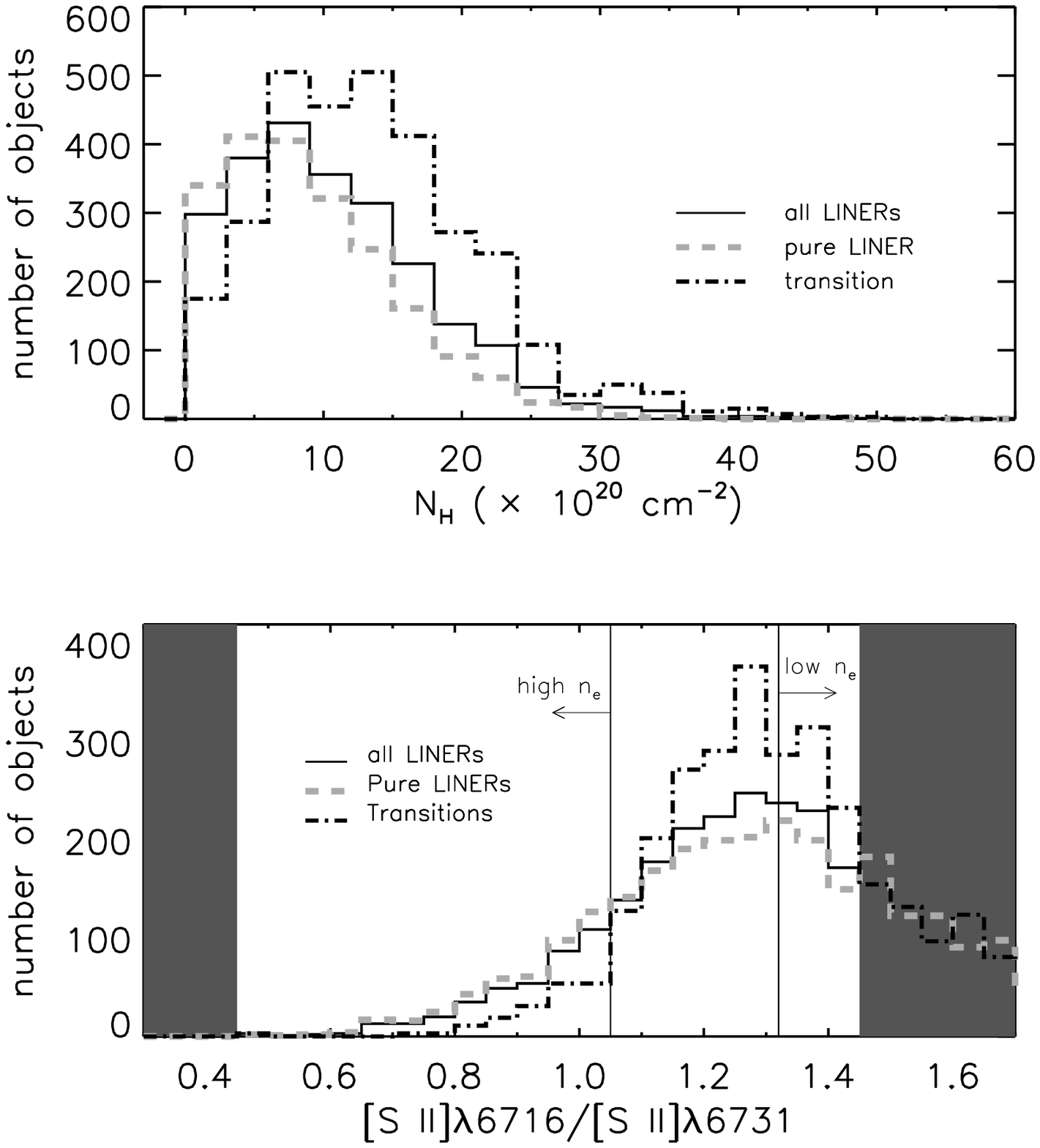}
\caption{ Comparison of distributions in $L_{\rm [O III]}$, $L_{\rm [O
I]}$, $N_H$, and $n_e$ for Transition objects and Pure LINERs.  The
histograms are re-normalized to have the total number of objects equal
to that of the whole LINER sample.  Transition object properties show
a slight tendency toward a more H {\sc ii}-like behavior.
\label{transition2}}
\end{figure}

We now turn to consider the Transition objects using similar
arguments.  Figure ~\ref{transition} illustrates the CFs estimated for
the Transition and Pure LINER samples (see definition in Table
~\ref{tbl-0}) along with a comparison of the distributions in $C$ and
the $u-r$ color for these samples.  Figure ~\ref{transition2}
illustrates how Transition systems and the Pure LINERs compare in
terms of $L_{\rm [O I]}$, $L_{\rm [O III]}$, $N_H$, and $n_e$.  The
results of the power-law fitting of the correlation functions, that
are also recorded in Table ~\ref{tbl-1}, show a great discrepancy in
the clustering amplitudes of these two sub-categories of LINERs.  The
Pure LINERs, which constitute the majority of the LINER sample, remain
strongly clustered with an amplitude similar to that measured for the
whole LINER sample and consistent with that of the whole galaxy
sample. The Transition objects are clearly less clustered than the
Pure LINERs, and are comparable in their clustering amplitude to
Seyferts and/or H {\sc ii}'s.

As expected, the intrinsic properties of the Transition objects
suggest a tendency toward a more Seyfert or H {\sc ii}-like behavior.
These systems show systematically larger $L_{\rm [O I]}$
and $L_{\rm [O III]}$, and higher $N_H$ than the Pure LINERs.
However, the nature of the Transition objects continues to remain
ambiguous. Their lower $n_e$ makes them more similar to the
star-forming systems than to Seyferts, while their host galaxies
remain in general of early-type, in the range of those that usually
harbor Seyferts, not H {\sc ii}'s.  Their clustering amplitude is
significantly lower than that of the typical red, early type galaxies.
Thus, a possibility for the power source in these objects is a small
black hole that is accreting at rates that are higher than those in
LINERs but still low, seemingly due to the general fuel deficiency.
Given that the obscuration remains relatively low in these sources,
moderate activity originating from such a source could explain their
intermediate behavior between that of traditional AGN and H {\sc ii}
systems.


Our own Milky Way seems to provide the most nearby example of a
typical H {\sc ii} or H {\sc ii}/LINER transition object, as outlined
above. There is compelling evidence for the presence of a relatively
low mass Galactic black hole \citep{scho03, ghe04} that is accreting,
as shown by radio (Sgr A$^*$ source, Balick \& Brown 1974) and X-ray
\citep{bag03} emission, and which exhibits large circumnuclear
extinction ($A_V \approx$ 31 mag; Rieke, Rieke \& Paul 1989). In terms
of its nebular luminosity, surface brightness distribution, and
ionization, \citet{shi05} reason that the Galaxy may be a typical
example of a H {\sc ii} or maybe a transition system, although its Sbc
Hubble type is most consistent with the former type of emission-line object.


\subsection{Environment \& AGN Activity}

The importance of the role that interactions with other galaxies plays
in triggering efficient gas fueling in AGN remains a matter of debate.
Although there is some evidence that the frequency of occurrence of
Seyferts in pairs and groups is higher than among isolated galaxies
\citep{kel98, kel04}, no clear relationships are found between the
presence of AGN and detailed morphological properties that are
signatures strong interactions (e.g., bars, see Keel 1996, Ho et
al. 1997).

A problem with many previous investigations of the environments of AGN
is that their definition as a class relies on relaxed definitions of
such objects which result, as illustrated in Section ~\ref{samples},
in samples dominated by LINERs. The influence of Seyfert-like
behavior remains negligible in such studies due to the small fraction of
such objects. For this reason, results that infer that AGN
environments are undistinguishable from those of the normal, inactive
galaxies \citep{mil03, wak04} apply for LINERs, but not necessarily
for Seyferts.  We find that LINERs and Seyferts differ not only in
their intrinsic physical characteristics (ionization level,
obscuration, density of their line-emitting gas) but also in their
clustering properties.  It is also important to understand that
environmental properties may vary with spatial scale.
For instance, the unusually steep slope of the Seyferts' CF
suggests that these sources may present extra clustering at
small scales, despite a generally low clustering amplitude at scales
$> 10 h^{-1}$ Mpc.  Thus, it is possible that LINERs and Seyferts
share similar small scale habitats, while their large scale
environments are quite different. 

To compare the small-scale environments of different classes of AGN,
we examine local galaxy density on a scale $\sim 1h^{-1}$Mpc.  For the
spectroscopic samples we built and analyzed in this study, we employ
~\citet{bla05} galaxy overdensities estimated by counting all galaxies
around each object in cylinders $1h^{-1}$ Mpc in radius and 1,600 km/s
long (in the redshift direction) centered on the source.  The tracer
galaxies are a volume-limited sample with $0.017 < z < 0.082$, and
$-23.5 <$ M$_{r^*} < -19.5$.  We find that the median overdensities
for Seyferts, LINERs and H {\sc ii}'s are 6.15, 6.13, and 4.17,
respectively, suggesting more crowded environments for
accretion-powered AGN than for objects with line emission dominated by
star formation.  The passive galaxies show the largest median
overdensity, 9.39.  For comparison, the median overdensity for the
whole galaxy sample is 6.34.  These results support our inferences
from estimates of the CF that suggest a higher density environment for
passive galaxies and a relatively lower local density around the H
{\sc ii}'s,.  The similarity in galaxy densities around Seyferts and
LINERs on this small scale may be a consequence of the Seyferts' extra
clustering at small scales; the density estimates characterize the
environment only at scales $\sim 1 h^{-1}$ Mpc, while the clustering
amplitudes, which show LINERs to be more strongly clustered, are
obtained by fitting the CFs at scales $3< r < 10 h^{-1}$ Mpc.

If the typical environments of LINERs are no different than those of
galaxies, then major mergers, harassment \citep{lak98}, or other
cluster-related mechanisms are unlikely to be the trigger for fueling
nuclear activity in these objects.  As discussed by
\citet{was05}, because the number of close companions of modest
luminosity AGN (i.e., LINERs) is very similar to that of non-active
galaxies \citep{was05, gro03}, and because their spatial distribution
is unbiased relative to that of the normal galaxy population (Miller et
al. 2003, Wake et al. 2004, Section ~\ref{class} of this study), a
possible cause of activity in these AGN may be {\it minor} mergers.

Seyferts, on the other hand, might still be caused by, e.g., galaxy
harassment, with these more luminous AGN being relatively recent
additions to clusters, and thus situated either at the periphery or in
the process of falling into dense regions.  A clumpy small-scale
spatial distribution would then be a natural consequence of such
environments.  Such migrants are also more likely to have large gas
reservoirs and only slightly disturbed morphologies.  We have shown
that Seyferts are distinct from other emission-line systems in their
high electron densities. So far, morphological studies of Seyfert
hosts reach no clear, statistically significant conclusion on the
degree of harassment in comparison with those of the normal galaxies:
Seyferts' hosts exhibit similar morphologies with those of field
galaxies \citep{derob98a, derob98b}, do not particularly show evidence
for bars \citep{reg99, lai02} or for galaxy-galaxy interactions
\citep{mal98}, while results of pair counting in both optical and IR,
remain inconclusive, some studies finding possible excess of
companions \citep{dah84, raf95}, while others not \citep{fue88, lau95,
sch01}.  Minor mergers that leave no or very little optical trace
\citep{cor00} might however explain fueling that would originate from
a canibalized secondary, less massive galaxy that thus causes only
minor perturbations to the primary galaxy.  Such events might as well
happen at the galaxy clusters' peripheries.

\section{Summary \& Conclusions} \label{summary}


We investigate the spatial clustering of low-luminosity AGN in the
nearby universe, using spectroscopic classification and measurement of
related physical parameters to examine this clustering as a function
of their nuclear emission properties.  We estimate and compare the CFs
for sub-samples selected on these various properties, and analyze
these results in light of the well-known dependence of clustering on
halo mass, as well as the empirical relation between black hole mass
and galaxy mass.  This study reveals a strong connection between the
galaxy clustering amplitudes, hence the properties of their dark
matter halos, and the intrinsic physical properties of these objects'
line-emitting regions. A brief overview of our main results
is given in Table ~\ref{tbl-summary}, and we list our
conclusions as follows.

(i) The amplitude of the two-point correlation function is strongly
dependent on galactic nuclear emission properties.  By employing
detailed spectroscopic classification schemes we find that, contrary
to previous claims, bona fide low luminosity AGN, spectroscopically
classified as Seyfert galaxies, are clearly less clustered than the
average galaxies, and therefore do not tracing the same
underlying structures.  LINERs, on the other hand, exhibit a
clustering amplitude higher than that of Seyferts, and consistent with
that of the whole galaxy sample.

(ii) AGN defined using only the [\ion{N}{2}]/H$\alpha$
vs. [\ion{O}{3}]/H$\beta$ diagnostic diagram are dominated by the
LINER-like objects, and include a significant fraction of objects with
conspicuously strong nuclear star-formation activity.  Two, or better,
three-dimensional diagnostic diagrams that employ forbidden lines
whose emissivity is less sensitive to abundance effects, like
[\ion{S}{2}] and/or [\ion{O}{1}], are necessary to
accurately separate sources powered by accretion from
those in which star-forming activity is dominant.

It is, therefore, important to emphasize that more lenient AGN
definitions are strongly biased toward LINER-like behavior, and
consequently, do not represent the traditional AGN activity and
physical characteristics.  LINERs differ from Seyferts in many of
their intrinsic characteristics.  In LINERs, the emitting gas density
is lower than in Seyferts, and the obscuration is generally at very
low levels.  As suggested by the large differences we observe in the
[\ion{O}{1}] and [\ion{O}{3}] line luminosities, the accretion
activity is significantly reduced in LINERs while in Seyferts is
generally high.  

(iii) Emission parameters such as $L_{\rm [O III]}$ and, in
particular, $L_{\rm [O I]}$, which are sensitive to the level of
nuclear accretion activity, play an important role in differentiating
systems of different clustering properties. Highly luminous and,
therefore, more active objects are significantly less clustered than
faint, less active ones; the latter types are strongly represented
by LINERs.

(iv) The amount and nature of the fuel available for nuclear activity,
whether accretion or star-formation, is also connected to the spatial
clustering characteristics.  The less obscured systems and the galaxies
with moderately high emitting gas densities are more clustered.  Such
sources are highly represented by LINERs.

(v) Host galaxy properties and, consequently, the morphology-density
relation, do not strongly influence the difference in the spatial
clustering found between Seyferts and LINERs, and play only a
secondary role in the large differences between the clustering
amplitudes of the low and high $L_{\rm [O I]}$ and/or $L_{\rm [O
III]}$ subsets.

(vi) Based on the strong relationship between host halo mass, as given
by the clustering amplitude, and the spectral properties of the
line-emitting galaxies, we speculate that: 1) Seyfert activity arises
from active accretion onto small BHs that is abundantly fueled, while
2) LINERs' emission-line properties originate from ionization by
slowly accreting massive BHs whose circumnuclear material supply is
generally low.
In this scenario, H {\sc ii} systems may be interpreted as harboring
relatively small, weakly accreting BHs whose activity remains
enshrouded within the surrounding star-forming activity.  The
similarity of the Milky Way nuclear system to that of a typical H {\sc
ii} galaxy in this picture provides a means of integrating the
Galactic Center into the larger context of galaxy nuclei.


\acknowledgements Support for this work was provided by NASA through
grant NAG5-12243. M. S. V. thanks the Departmental of Astrophysical
Sciences at Princeton University and the Korea Institute for Advanced
Study for their hospitality and support. The authors thank David
Goldberg, Fiona Hoyle, and Joe Shields for helpful comments and
discussion.

    Funding for the SDSS and SDSS-II has been provided by the Alfred
    P. Sloan Foundation, the Participating Institutions, the National
    Science Foundation, the U.S. Department of Energy, the National
    Aeronautics and Space Administration, the Japanese Monbukagakusho,
    the Max Planck Society, and the Higher Education Funding Council
    for England. The SDSS Web Site is http://www.sdss.org/.

    The SDSS is managed by the Astrophysical Research Consortium for
    the Participating Institutions. The Participating Institutions are
    the American Museum of Natural History, Astrophysical Institute
    Potsdam, University of Basel, Cambridge University, Case Western
    Reserve University, University of Chicago, Drexel University,
    Fermilab, the Institute for Advanced Study, the Japan
    Participation Group, Johns Hopkins University, the Joint Institute
    for Nuclear Astrophysics, the Kavli Institute for Particle
    Astrophysics and Cosmology, the Korean Scientist Group, the
    Chinese Academy of Sciences (LAMOST), Los Alamos National
    Laboratory, the Max-Planck-Institute for Astronomy (MPA), the
    Max-Planck-Institute for Astrophysics (MPIA), New Mexico State
    University, Ohio State University, University of Pittsburgh,
    University of Portsmouth, Princeton University, the United States
    Naval Observatory, and the University of Washington.

\clearpage

\begin{deluxetable}{lrcccc}
\tablecolumns{6} \tablewidth{0pt} 
\tablecaption{Object Sample Definition: Spectral classes.
\label{tbl-0}} 
\tablehead{ 
\colhead{Sample} & 
\colhead{$N_{obj}$} &
\colhead{[\ion{O}{3}]/H$\beta$} &
\colhead{[\ion{N}{2}]/H$\alpha$} &
\colhead{[\ion{S}{2}]/H$\alpha$} &
\colhead{[\ion{O}{1}]/H$\alpha$}\\
\colhead{(1)} &
\colhead{(2)} &
\colhead{(3)} &
\colhead{(4)} &
\colhead{(5)} &
\colhead{(6)}} 
\startdata
Emission-line  & 54,410& $> 0$   & $> 0$     & $> 0$     &$> 0$   \\    
Seyferts       &  2,392& $\geq 3$& $\geq 0.6$& $\geq 0.4$&$\geq 0.08$   \\
LINERs         &  7,468& $< 3$   & $\geq 0.6$& $\geq 0.4$&$\geq 0.08$   \\  
...pure liners &  5,313& $< 3$   & $\geq 0.6$& $\geq 0.4$&$\geq 0.17$   \\  
...transition  &  2,155& $< 3$   & $\geq 0.6$& $\geq 0.4$&$\geq 0.08$,$< 0.17$\\  
H II's         & 34,781& $> 0$   & $< 0.6$   & $< 0.4$   &$< 0.08$   \\
\enddata 
\tablecomments{Columns (3) to (6) indicate the range of values in the 
line flux ratios.}
\end{deluxetable}

\begin{deluxetable}{lrccc}
\tablecolumns{7} \tablewidth{0pt} \tablecaption{Absolute Magnitude
Limited Correlation Function Samples
\label{tbl-1}} \tablehead{ 
\colhead{Sample} & 
\colhead{$N_{obj}$} &
\colhead{$s_0$} & 
\colhead{$\gamma$} &
\colhead{$\sigma_{s_0\gamma}/\sqrt{\sigma_{s_0} \sigma_{\gamma}}$} }
\startdata
all galaxies   &45,303 & 7.80$\pm$0.49 & 1.17$\pm$0.05 & --0.82 \\
passive        & 6,383 &11.42$\pm$1.17 & 1.28$\pm$0.07 & --0.82 \\
\hline
Emission-line\tablenotemark{a}&12,676& 6.33$\pm$0.94 & 1.30$\pm$0.16 & --0.15\\
Seyferts\tablenotemark{a}     &   829& 5.67$\pm$0.62 & 1.56$\pm$0.17 & --0.58\\
LINERs\tablenotemark{a}       & 3,263& 7.82$\pm$0.64 & 1.39$\pm$0.09 & --0.53\\
...pure LINERs                & 2,424& 7.39$\pm$0.67 & 1.33$\pm$0.11 & --0.46\\
...transition                 &   839& 5.38$\pm$0.71 & 1.35$\pm$0.23 & --0.18\\   
H II's\tablenotemark{a}       & 6,622& 5.81$\pm$0.53 & 1.28$\pm$0.11 &0.08\\
\hline
Emission-line\tablenotemark{b}&21,497 & 6.45$\pm$0.24 & 1.24$\pm$0.04 & --0.67 \\
Seyferts\tablenotemark{b}     & 1,018 & 6.00$\pm$0.64 & 1.41$\pm$0.15 & --0.48 \\
LINERs\tablenotemark{b}       & 5,326 & 7.26$\pm$0.61 & 1.29$\pm$0.08 & --0.73 \\
H II's\tablenotemark{b}       &12,649 & 5.73$\pm$0.24 & 1.22$\pm$0.06 & --0.46 \\
\hline
AGN (Kauffmann et al.)   &14,421& 7.44$\pm$0.37 & 1.25$\pm$0.07 & --0.44\\
non-AGN (Kauffmann et al.) &13,630& 6.26$\pm$0.69 & 1.24$\pm$0.11 & --0.52\\
AGN (Miller et al.)       & 9,846& 7.01$\pm$0.43 & 1.27$\pm$0.06 & --0.63\\
non-AGN (Miller et al.)   &15,178& 5.62$\pm$0.59 & 1.24$\pm$0.12 & 0.03\\
\hline
log $L_{\rm [O III]}$\tablenotemark{c} $>$ 40    & 2,041 & 6.22$\pm$0.57 & 1.41$\pm$0.11 & --0.24 \\
38 $<$ log $L_{\rm [O III]}$\tablenotemark{c} $<$ 39&10,464 & 7.17$\pm$0.44 & 1.21$\pm$0.05 & --0.63 \\
log $L_{\rm [O III]}$\tablenotemark{c} $<$ 37    & 1,580 & 8.79$\pm$0.75 & 1.31$\pm$0.10 & --0.54 \\
\hline
log $L_{\rm [O I]}$\tablenotemark{c} $>$ 39    & 4,561 & 6.11$\pm$0.27 & 1.34$\pm$0.07 & --0.36 \\
log $L_{\rm [O I]}$\tablenotemark{c} $<$ 38    & 2,153 &13.15$\pm$1.09 & 1.33$\pm$0.07 & --0.77 \\
\hline
high $N_H$& 2,575& 6.91$\pm$0.57 & 1.25$\pm$0.09 &--0.19\\
low $N_H$ & 2,921& 7.38$\pm$0.51 & 1.35$\pm$0.07 &--0.68\\
\hline
high $n_e$&   788& 7.50$\pm$0.98 & 1.15$\pm$0.13&--0.98 \\
low $n_e$ & 3,435& 5.51$\pm$0.45 & 1.43$\pm$0.11&--0.11 \\
\enddata 								     
\tablenotetext{a}{definition criteria include [\ion{O}{1}], [\ion{O}{1}]/H$\alpha$}
\tablenotetext{b}{definition criteria exclude [\ion{O}{1}], [\ion{O}{1}]/H$\alpha$}
\tablenotetext{c}{the line luminosities are in units of erg s$^{-1}$}
\tablecomments{All samples use $14.5 < r < 17.7$, $0.05 < z		     
  < 0.12$, and $ -21.6 < M_r < -20.2$. $s_0$ is in units of
  $h^{-1}$ Mpc.  $\sigma_{s_0\gamma}/\sqrt{\sigma_{s_0}
    \sigma_{\gamma}}$ is the normalized correlation coefficient 
between $s_0$ and $\gamma$.}
\end{deluxetable}

\begin{deluxetable}{lrcc}
\tablecolumns{4} \tablewidth{0pt} 
\tablecaption{Object Sample Definition as a function of $n_e$
\label{tbl-ne}} 
\tablehead{ 
\colhead{Sample} & 
\colhead{$N_{obj}$} &
\colhead{[\ion{S}{2}]$\lambda6716/\lambda$6731} &
\colhead{$n_e$(cm$^{-3}$)} }
\startdata
high $n_e$ &  5,922& $\leq 1.05$& $\ga 5\times 10^2$ \\  
low $n_e$  & 27,116& $> 1.32$   & $\la 1 \times10^2$ \\
\enddata 
\end{deluxetable}

\begin{deluxetable}{lrcc}
\tablecolumns{4} 
\tablewidth{0pt} 
\tablecaption{Object Sample Definition as a function of $N_H$
\label{tbl-nh}} 
\tablehead{ 
\colhead{Sample} & 
\colhead{$N_{obj}$} &
\colhead{$N_H$} &
\colhead{$(E-B)_V$} }
\startdata
low $N_H$  &19,343 & $< 10 \times10^{20}$ cm$^{-2}$& $< 0.20$ \\  
high $N_H$ & 8,573 & $> 20 \times10^{20}$ cm$^{-2}$& $> 0.45$ \\
\enddata 
\end{deluxetable}

\begin{deluxetable}{lrcccccc}
\tablecolumns{5} \tablewidth{0pt} \tablecaption{Median Values in
Absolute Magnitude Limited Samples
\label{tbl-param}} 
\tablehead{ 
\colhead{Sample} & 
\colhead{$N_{obj}$} &
\colhead{$C$} &
\colhead{$u - r$} &
\colhead{log $L_{\rm [O I]}$ } &
\colhead{log $L_{\rm [O III]}$ } &
\colhead{$N_H$} &
\colhead{$n_e$} \\
\colhead{(1)} &
\colhead{(2)} &
\colhead{(3)} &
\colhead{(4)} &
\colhead{(5)} &
\colhead{(6)} &
\colhead{(7)} &
\colhead{(8)}}
\startdata
emission-line  &12,676&2.61 &2.31&38.80 &39.26 &13.44 & 1.7 $\times 10^1$\\  
seyferts       &   829&2.82 &2.63&38.93 &39.81 &11.65 & 1.9 $\times 10^2$\\
liners         & 3,263&2.93 &2.73&38.34 &38.65 & 9.64 & 1.2 $\times 10^2$\\  
...pure LINERs & 2,424&2.99 &2.79&38.22 &38.48 & 8.19 & 1.2 $\times 10^2$\\
...transition  &   839&2.73 &2.53&38.70 &39.15 &12.81 & 1.0 $\times 10^2$\\
H II's         & 6,622&2.38 &2.01&38.90 &39.37 &15.92 & 1.6 $\times 10^1$\\  
\enddata 				  
\tablecomments{ Col. (2): number of objects in the 
samples; Col. (3): concentration index; Col. (4): $u-r$ color
for the host galaxies; Cols. (5) and (6): logarithmic values of 
[\ion{O}{1}] and [\ion{O}{3}] luminosities in erg s$^{-1}$; Col. (7): neutral
hydrogen column density in units of $10^{20}$ cm$^{-2}$; Col. (8):  
electron density in cm$^{-3}$}
\end{deluxetable}

%

\begin{deluxetable}{l|c|c|c|c}
\tablecolumns{5} \tablewidth{0pt} 
\tablecaption{Summary of Results \& Interpretation
\label{tbl-summary}} 
\tablehead{ 
\colhead{ {$_{\rm Properties}\diagdown^{\rm Sample}$}} & 
\colhead{H {\sc ii}} &
\colhead{Transition} &
\colhead{Seyferts} &
\colhead{Pure LINERs}}
\startdata
clustering    &weak     &weak        & weak         & strong\\
~~~~($s_0$)  &         &            &              &{\it -like galaxies}\\  
\hline
fuelling rate &average&moderately low&relatively high& (very) low\\
($L_{\rm [O I]}$, $L_{\rm [O III]}$)    
              &({\it not low!}) &  &{\it efficient?} &{\it inefficient?}\\
\hline
fuel available & low   & low          & high         & moderately high\\  
 ~~~~($n_e$)   &       &              &              & (wide range)\\  
\hline
obscuration   & high  & low          & wide range   & low\\  
\hline
host-         &blue-ish &red-ish     &red-ish       & red-ish\\  
-morphology   &late type&earlier type&earlier type  &earlier type\\  
              &         &{\it -like Seyferts}&      & {\it -like Seyferts}\\  
\hline
$M_{\rm BH}$  & small   & small      & small              & large\\  
\hline
life-time     & ?       & ?          & short              & long \\  
\enddata 
\end{deluxetable}

\clearpage

\clearpage

\clearpage

\clearpage

\clearpage

\clearpage

\clearpage

\clearpage

\clearpage

\clearpage

\end{document}